\pdfoutput=1
\documentclass[11pt,twoside,a4paper,cmspaper,final,collab]{cms-tdr}
\def\svnVersion{8159592}\def\svnDate{2021/11/26}\def\cmsCernNoTag{CERN-EP-2021-117}\def\cmsCernDate{\today}\def\cmsMessage{Published in the Journal of High Energy Physics as \href{http://dx.doi.org/10.1007/JHEP12(2021)180}{\doi{10.1007/JHEP12(2021)180}.}}
\begin{document}\cmsNoteHeader{TOP-18-010}

\newlength\cmsFigWidth
\newlength\cmsTabSkip\setlength{\cmsTabSkip}{1ex}
\ifthenelse{\boolean{cms@external}}{\setlength\cmsFigWidth{0.49\textwidth}}{\setlength\cmsFigWidth{0.65\textwidth}}
\ifthenelse{\boolean{cms@external}}{\providecommand{\cmsLeft}{upper\xspace}}{\providecommand{\cmsLeft}{left\xspace}}
\ifthenelse{\boolean{cms@external}}{\providecommand{\cmsRight}{lower\xspace}}{\providecommand{\cmsRight}{right\xspace}}

\newcommand{\pp}{\ensuremath{\Pp\Pp}\xspace}
\newcommand{\fb} {\mbox{\ensuremath{\,\text{fb}}}\xspace}
\newcommand{\pb} {\mbox{\ensuremath{\,\text{pb}}}\xspace}
\newcommand{\HERWIGSeven} {{\HERWIG{}7}\xspace}
\newcommand{\ttg}{\ensuremath{\ttbar\gamma}\xspace}
\newcommand{\ttgVert}{\ttg}
\newcommand{\ttZ}{\ensuremath{\ttbar\PZ}\xspace}
\newcommand{\ttW}{\ensuremath{\ttbar\PW}\xspace}
\newcommand{\tZq}{\ensuremath{\PQt\PZ\Pq}\xspace}
\newcommand{\tW}{\ensuremath{\PQt\PW}\xspace}
\newcommand{\VV}{\ensuremath{\PV\PV}\xspace}
\newcommand{\WW}{\ensuremath{\PW\PW}\xspace}
\newcommand{\WZ}{\ensuremath{\PW\PZ}\xspace}
\newcommand{\ZZ}{\ensuremath{\PZ\PZ}\xspace}
\newcommand{\tg}{\ensuremath{\PQt\gamma}\xspace}
\newcommand{\WJets}{\ensuremath{\PW\text{+jets}}\xspace}
\newcommand{\WGamma}{\ensuremath{\PW\gamma}\xspace}
\newcommand{\tWGamma}{\ensuremath{\PQt\PW\gamma}\xspace}
\newcommand{\ZGamma}{\ensuremath{\PZ\gamma}\xspace}
\newcommand{\Mthree}{\ensuremath{M_3}\xspace}
\newcommand{\mtw}{\ensuremath{\mT(\PW)}\xspace}
\newcommand{\mlg}{\ensuremath{m(\ell,\gamma)}\xspace}
\newcommand{\ptG}{\ensuremath{\pt(\gamma)}\xspace}
\newcommand{\abseta}{\ensuremath{\abs{\eta}}\xspace}
\newcommand{\etaG} {\ensuremath{\abs{\eta(\gamma)}}\xspace}
\newcommand{\dRlg}{\ensuremath{\Delta R(\ell,\gamma)}\xspace}

\newcommand{\muR}{\ensuremath{\mu_\mathrm{R}}\xspace}
\newcommand{\muF}{\ensuremath{\mu_\mathrm{F}}\xspace}

\newcommand{\nLep}{\ensuremath{N_\ell}}
\newcommand{\nG}{\ensuremath{\ensuremath{N_\gamma}}\xspace}
\newcommand{\nJet}{\ensuremath{N_\text{j}}\xspace}
\newcommand{\nBTag}{\ensuremath{N_\PQb}\xspace}
\newcommand{\relIso}{\ensuremath{I_\textrm{rel}}\xspace}
\newcommand{\chIso}{\ensuremath{I_\textrm{chg}(\gamma)}\xspace}
\newcommand{\sieie}{\ensuremath{\sigma_{\eta\eta}(\gamma)}\xspace}

\newcommand{\ctZ}{\ensuremath{c_{\PQt\PZ}}\xspace}
\newcommand{\ctZI}{\ensuremath{c_{\PQt\PZ}^\mathrm{I}}\xspace}
\newcommand{\ctA}{\ensuremath{c_{\PQt\gamma}}\xspace}
\newcommand{\ctAI}{\ensuremath{c_{\PQt\gamma}^\mathrm{I}}\xspace}

\newcommand{\thetaw}{\ensuremath{\theta_{\PW}}}
\newcommand{\sinw}{\ensuremath{\sin\thetaw}}
\newcommand{\cosw}{\ensuremath{\cos\thetaw}}

\cmsNoteHeader{TOP-18-010}
\title{Measurement of the inclusive and differential \texorpdfstring{$\ttbar\gamma$}{ttgamma} cross sections in the single-lepton channel and EFT interpretation at \texorpdfstring{$\sqrt{s}=13\TeV$}{sqrt(s)=13 TeV}}

\date{\today}

\abstract{
The production cross section of a top quark pair in association with a photon is measured in proton-proton collisions at a center-of-mass energy of 13\TeV.
The data set, corresponding to an integrated luminosity of 137\fbinv, was recorded by the CMS experiment during the 2016--2018 data taking of the LHC.
The measurements are performed in a fiducial volume defined at the particle level.
Events with an isolated, highly energetic lepton, at least three jets from the hadronization of quarks, among which at least one is \PQb~tagged, and one isolated photon are selected.
The inclusive fiducial \ttg cross section, for a photon with transverse momentum greater than 20\GeV and pseudorapidity $\abseta<1.4442$, is measured to be $798\pm 7\stat\pm 48\syst\fb$, in good agreement with the prediction from the standard model at next-to-leading order in quantum chromodynamics.
The differential cross sections are also measured as a function of several kinematic observables and interpreted in the framework of the standard model effective field theory~(EFT), 
leading to the most stringent direct limits to date on anomalous electromagnetic dipole moment interactions of the top quark and the photon.
}

\hypersetup{%
pdfauthor={CMS Collaboration},%
pdftitle={Measurement of the inclusive and differential ttgamma cross sections in the single-lepton channel and EFT interpretation at sqrt(s) = 13 TeV},%
pdfsubject={CMS},%
pdfkeywords={CMS,  top quark}}

\maketitle

\section{Introduction}\label{sec:intro}

The large amount of proton-proton (\pp) collision data at a center-of-mass energy of 13\TeV at the LHC allows for precision measurements of standard model (SM) processes with small production rates.
Among these, top quark production provides a testing ground for the SM predictions and for phenomena beyond the SM (BSM). 
In particular, precise measurements of the inclusive and differential cross sections of top quark pair production in association with a high-energy photon (\ttg) constrain anomalous \ttgVert electroweak interactions~\cite{Baur:2004uw,Bouzas:2012av,Schulze:2016qas,Rontsch:2015una}.

The CDF Collaboration at the Fermilab Tevatron measured the \ttg production cross section using proton-antiproton~collisions 
at $\sqrt{s} = 1.96\TeV$~\cite{CDFttgamma}, while at the LHC the measurement was performed in 
\pp~collisions at 7\TeV by the ATLAS~\cite{ATLASttgamma7TeV}, and at 
8\TeV by both the ATLAS~\cite{ATLASttgamma8TeV} and CMS~\cite{CMSttgamma8TeV} Collaborations.
At 13\TeV, the ATLAS Collaboration measured inclusive and differential \ttg production cross sections in leptonic~\cite{Aaboud:2018hip} and in the $\Pe\Pgm$~\cite{Aad:2020axn} final states.
All of these results are in agreement with the SM.

\begin{figure}[hb]
    \centering
    \includegraphics[width=0.43\textwidth]{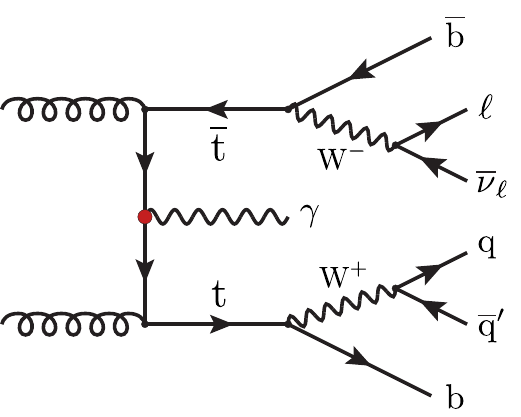}\hfil
    \includegraphics[width=0.40\textwidth]{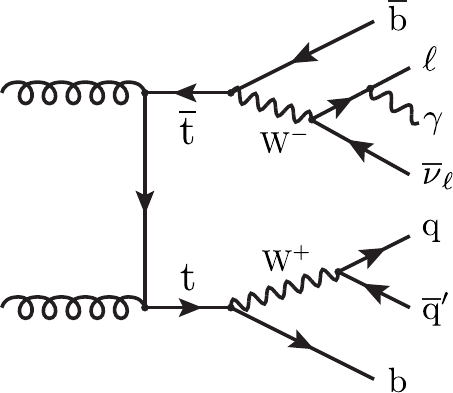}
    \caption{Representative LO Feynman diagrams for the \ttg signal process in the single-lepton channel, where the highly energetic photon originates from the top quark (left), or is emitted from a lepton (right). The \ttgVert interaction vertex is indicated by a circle.}
    \label{fig:FeynDiagram}
\end{figure}

In this paper, the inclusive and differential \ttg production cross sections are measured in \pp~collisions at $\sqrt{s}=13\TeV$.
The analysis uses a data sample recorded with the CMS detector during Run~2~(2016--2018) of the LHC, which corresponds to an integrated luminosity of 137\fbinv.
The measurement is performed in the single-lepton (electron or muon) final state in a fiducial region defined at particle level.
The inclusive fiducial \ttg cross section is measured for a selection on the photon transverse momentum of $\ptG>20\GeV$ and the pseudorapidity of $\etaG<1.4442$,
corresponding to the barrel region of the CMS electromagnetic calorimeter (ECAL).
Differential cross sections are measured in the same fiducial region as a function of \ptG, \etaG, and the angular separation between the lepton and the photon, \dRlg.
The observations are interpreted in the context of the SM effective field theory (SM-EFT)~\cite{Bylund:2016phk}, where the \ctZ and \ctZI operators, defined in Ref.~\cite{AguilarSaavedra:2018nen}, are constrained using the measurement of the distribution of \ptG.
Tabulated results are provided in HEPDATA~\cite{hepdata}.
Examples of Feynman diagrams at leading order (LO) contributing to the \ttg signal topology are shown in Fig.~\ref{fig:FeynDiagram}.

This paper is organized as follows. 
The CMS detector is briefly introduced in Section~\ref{sec:CMS-detector}. 
Details on the simulation of the signal and background processes and their modeling are provided in Section~\ref{sec:simulation}.
The online selection, event reconstruction, and object definitions are described in Section~\ref{sec:reco}.
The fiducial phase space definition and photon categorization are described in Section~\ref{sec:sel_fiducial}.
The event selection and the statistical treatment are discussed in Section~\ref{sec:ana_strategy}.
The procedures to estimate the backgrounds are described in Section~\ref{sec:bkg} and the systematic uncertainties are discussed in Section~\ref{sec:systematics}.
The obtained results and the interpretation of the measurements in the context of SM-EFT are presented in Section~\ref{sec:results}.
Finally, a summary is provided in Section~\ref{sec:summary}.

\section{The CMS detector}\label{sec:CMS-detector}

The central feature of the CMS apparatus is a superconducting solenoid of 6\unit{m} internal diameter, providing a magnetic field of 3.8\unit{T}.
Within the solenoid volume are a silicon pixel and strip tracker, a lead tungsten crystal ECAL, and a brass and scintillator hadron calorimeter (HCAL), each composed of a barrel and two endcap sections.
Forward calorimeters extend the $\eta$~coverage provided by the barrel and endcap detectors that improve the measurement of the imbalance in transverse momentum.
Muons are detected in gas-ionization chambers embedded in the steel flux-return yoke outside the solenoid.

Events of interest are selected using a two-tiered trigger system~\cite{Khachatryan:2016bia}.
The first level trigger~(L1)~\cite{Sirunyan:2020zal}, composed of custom hardware processors, uses information from the calorimeters and muon detectors to 
select events at a rate of around 100\unit{kHz} within a fixed latency of about 4\mus. 
The second level, known as the high-level trigger, consists of a farm of processors running a version of the full event reconstruction software optimized for fast processing, 
and reduces the event rate to around 1\unit{kHz} before data storage~\cite{Khachatryan:2016bia}. 
A more detailed description of the CMS detector, together with a definition of the coordinate system used and the relevant kinematic variables, can be found in Ref.~\cite{Chatrchyan:2008zzk}.

\section{Simulated event samples}\label{sec:simulation}

{\tolerance=800 Multiple Monte Carlo (MC) event generators are used to simulate the background and signal contributions, matching the varying conditions for each data-taking period. 
The \ttbar, $t$-channel single top quark, \tW, and \WW background processes are simulated at next-to-LO~(NLO) in perturbative quantum chromodynamics~(QCD) with the \POWHEG~v2~\cite{Nason:2004rx,Frixione:2007vw,Alioli:2010xd,Campbell:2014kua,Alioli:2009je,Frixione:2007nw,Re:2010bp} event generator.
The QCD multijet processes are generated with \PYTHIA~v8.226 (8.230)~\cite{Sjostrand:2014zea} for the 2016 (2017, 2018) data-taking period.
All other background processes are simulated with \MGvATNLO~v2.6.0~\cite{Alwall:2014hca} at LO or NLO accuracy.
The \ttbar\ simulation is normalized to a cross section of $832\pm 42\pb$ calculated with the \textsc{Top++} v2.0 program~\cite{Czakon:2011xx}
at next-to-NLO (NNLO), including resummation of next-to-next-to-leading-logarithm (NNLL) soft-gluon terms~\cite{Beneke:2011mq,Cacciari:2011hy,Baernreuther:2012ws,Czakon:2012zr,Czakon:2012pz,Czakon:2013goa}.
Events with an $s$- or $t$-channel produced top quark are normalized to NLO cross sections~\cite{Aliev:2010zk,Kant:2014oha}, while the normalizations of \WW and \tW are at NNLO~\cite{Kidonakis:2015nna}.
The overlap of the \tW and \ttbar simulation is removed using the diagram removal technique~\cite{Frixione:2008yi}. 
Drell--Yan and \WJets events are generated with up to four extra partons in the matrix element calculations with \MGvATNLO at LO and are normalized to NNLO cross sections~\cite{Melnikov:2006kv,Catani:2009sm,Anastasiou:2003ds} including  electroweak corrections at NLO~\cite{Dittmaier:2014qza,Lindert:2017olm}.
The \WZ, \ZZ, \ZGamma, and \ttW samples are simulated at NLO precision with one extra parton at ME level.
The \WGamma sample is simulated at LO precision with up to three extra partons.
The \ttZ, \ttW, \tZq, \tg, \WGamma, \ZGamma, and other diboson processes~(\WZ or \ZZ) are normalized to the most precise cross sections available~\cite{Garzelli:2012bn,Frixione:2015zaa}.\par}

The \ttg signal is generated with \MGvATNLO~v2.6.0 at LO as a doubly resonant $2\to 7$ process including all decay channels of the intermediate \PW bosons.
It includes events where the photon is radiated from the intermediate top quarks, the intermediate \PW bosons and their decay products, the \PQb quarks, and, in the case of quark-antiquark annihilation, radiation from initial-state quarks. 
The photon is required to satisfy $\pt>10\GeV$ and $\abseta<5$, while the lepton must pass $\abseta<5$. 
The angular separation \DR~between the photon and any of the seven final-state particles is required to be greater than 0.1, where
$\DR=\sqrt{\smash[b]{(\Delta\eta)}^2+\smash[b]{(\Delta\phi)}^2}$ and $\phi$ denotes the azimuthal angle.
The renormalization scale (\muR) and factorization scale (\muF) are set to $\frac{1}{2}\sum_i \sqrt{\smash[b]{m_i}^2+\smash[b]{p_{\mathrm{T},i}}^2}$, where the sum runs over all final-state particles generated at the matrix-element~(ME) level.
Although no photons are simulated at the ME level in the \ttbar process, initial- and final-state photon radiation is accounted for in the showering algorithm.
We remove double counting of the \ttbar and \ttg samples by excluding events from the \ttbar sample with a generated photon passing the photon requirements of the \ttg signal sample.
The overlap between \WGamma and \WJets, \ZGamma and Drell--Yan, and \tg and the single top quark $t$-channel process is removed analogously.

{\renewcommand{\arraystretch}{1.1}
\begin{table*}[t]
\centering
\topcaption{
Event generator, perturbative order in QCD of the simulation, and perturbative order of the cross section normalization for each process.
}
\label{table:samples}
\begin{tabular}{cccl}
\multirow{2}{*}{Process}   & \multirow{2}{*}{Event generator} & Perturbative          & Cross section       \\
                           &                                  & order of simulation  & normalization       \\
\hline\ttg                 & \MGvATNLO                        & LO &NLO\\
\ttbar                     & \POWHEG & NLO& NNLO+NNLL~\cite{Beneke:2011mq,Cacciari:2011hy,Baernreuther:2012ws,Czakon:2012zr,Czakon:2012pz,Czakon:2013goa,Czakon:2011xx} \\
Single~\PQt~($t$-channel)        & \POWHEG              & NLO & NLO~\cite{Aliev:2010zk,Kant:2014oha}             \\
Single~\PQt~($s$-channel)        & \MGvATNLO            & NLO & NLO~\cite{Aliev:2010zk,Kant:2014oha}           \\
\tW                               & \POWHEG              & NLO & NNLO~\cite{Kidonakis:2015nna}                              \\
Drell--Yan, \WJets         & \MGvATNLO& LO& NNLO~\cite{Melnikov:2006kv,Catani:2009sm,Anastasiou:2003ds,Dittmaier:2014qza,Lindert:2017olm}\\ 
\WGamma                     & \MGvATNLO                        & LO& NLO               \\
\WW                        & \POWHEG & NLO& NNLO~\cite{Gehrmann:2014fva} \\
\tg, \ZGamma, \WZ, \ZZ     & \multirow{2}{*}{\MGvATNLO} & \multirow{2}{*}{NLO} & \multirow{2}{*}{NLO} \\
\ttZ, \ttW, \tZq           &                & & \\
Multijet                   & \PYTHIA & LO & LO 
\end{tabular}
\end{table*}
}

The event generators are interfaced with \PYTHIA v8.226 (8.230) using the CP5 tune~\cite{Skands:2014pea,Khachatryan:2015pea,Sirunyan:2019dfx} for the 2016 (2017, 2018) samples to simulate multiparton interactions, fragmentation, parton shower, and hadronization of partons in the initial and final states, along with the underlying event.
The NNPDF parton distribution functions (PDFs)~v3.1~\cite{Ball:2017nwa} are used according to the  different perturbative order in QCD at the ME level.
For the 2016 data-taking period, the CUETP8M1 tune~\cite{Khachatryan:2015pea} and the NNPDF PDFs~v3.0~\cite{Ball:2014uwa} are used for the Drell--Yan, \WJets, \tg, \ZGamma, \WGamma, diboson, \ttW, \ttZ, \tZq, and multijet processes.
Double counting of the partons generated with \MGvATNLO and \PYTHIA is removed using the MLM~\cite{Alwall:2007fs} and the \textsc{FxFx}~\cite{Frederix:2012ps} matching schemes for LO and NLO samples, respectively.
The events are subsequently processed with a \GEANTfour-based simulation model~\cite{Agostinelli:2002hh} of the CMS detector.
All simulated samples include the effects of additional \pp collisions in the same or adjacent bunch crossings (pileup), and are reweighted according to the observed distribution of the number of interactions in each bunch crossing~\cite{Sirunyan:2018nqx}.
In the following, to simplify the notation, the single top quark, \ttbar, and \tg processes are grouped in the $\PQt/\ttbar$ category, and furthermore, the \tZq, \ttW, \ttZ, \WW, \WZ, and \ZZ processes in a category labeled ``other''. 
A summary of the event samples is provided in Table~\ref{table:samples}.

\section{Event reconstruction}\label{sec:reco}

Events are selected at the high-level trigger by the algorithms that require the presence of at least one lepton ($\ell=\Pe$ or \Pgm).
The trigger threshold on the leading muon \pt is 27 (24)\GeV in the 2017 (2016 and 2018) LHC running period.
For electrons, the trigger threshold in the 2016 (2017--2018) period is 27 (32)\GeV.

The particle-flow~(PF) algorithm~\cite{PFpaper} aims to reconstruct and identify each individual particle in an event, with an optimized combination of information from the various elements of the CMS detector.
The candidate vertex with the largest value of summed physics-object $\pt^2$ is taken to be the primary \pp interaction vertex (PV).
The energy of charged PF hadrons is determined from a combination of the track momentum and the matching ECAL and HCAL energy deposits, corrected for zero-suppression effects and for the response function of the calorimeters to hadronic showers. 
The energy of neutral PF hadrons is obtained from the corresponding corrected ECAL and HCAL energies.

The energy of electrons is determined from a combination of the electron momentum at the PV as determined by the tracker,
the energy of the corresponding ECAL cluster, and the energy sum of all bremsstrahlung photons spatially compatible with originating from the electron track.
Electron candidates are required to satisfy $\pt>35\GeV$ and $\abseta<2.4$, excluding the transition region between the barrel and endcap of the ECAL, $1.4442<\abseta<1.5660$.
The electron identification is performed using shower shape variables, track-cluster matching variables, and track quality variables. 
To reject electrons originating from photon conversion inside the detector, electrons are required to miss at most one possible hit in the innermost tracker layer and to be incompatible with any conversion-like secondary vertices. 

The momentum of muons is obtained from the curvature of the corresponding track. 
Muon candidates are selected having $\pt>30\GeV$ and $\abseta<2.4$. 
The identification of muon candidates is performed using the quality of the geometrical matching between the measurements of the tracker and muon system~\cite{Sirunyan:2018fpa}.

The energy of photons is obtained from the ECAL measurement.
Photon candidates are required to satisfy $\ptG>20\GeV$ and to fall within the barrel of the ECAL,
$\abseta<1.4442$. 
The identification of photons is based on isolation and shower shape information as a function of \pt and $\eta$, and takes into account pileup effects~\cite{Khachatryan:2015hwa, Khachatryan:2015iwa}. 
In particular, the lateral shower extension must satisfy $\sieie<0.01$ for the chosen ``medium'' photon working point.
It is defined as the second moment of a log-weighted distribution of crystal energies in $\eta$, calculated in the $5\times5$ matrix around the most energetic crystal in the photon's supercluster~\cite{Sirunyan:2020ycc}.
Because of the reduced power of the \sieie observable in the ECAL endcap region in rejecting nonprompt photons, we find that excluding this $\abseta$ range improves the uncertainties in the measurements.

All lepton and photon candidates are required to be isolated from other objects by selecting the reconstructed charged and neutral PF candidates in a cone around the candidate.
A radius $\DR=0.3~(0.4)$ is used for electron (muon) candidates. 
For electron candidates, \pt- and $\eta$-dependent thresholds are set on the pileup corrected scalar \pt sum of photons and neutral and charged hadrons reconstructed by the PF algorithm~($\relIso(\Pe)$) in the range of 5--10\%.
The chosen ``tight'' electron working point has a 70\% efficiency while rejecting electron candidates originating from jets~\cite{Sirunyan:2020ycc}. 
A muon candidate is isolated if it satisfies $\relIso(\Pgm)<0.15$. 
The efficiency of the chosen ``tight'' working point is 90--95\%, depending on \pt and $\eta$ of the muon candidate~\cite{Sirunyan:2018fpa}. 
For photon candidates, the scalar \pt sum of the charged particles within a cone of $\DR=0.3$, denoted as the photon charged-hadron isolation, must satisfy $\chIso\leq1.141\GeV$.
Depending on the photon candidate \pt, there are separate requirements on the photon neutral-hadron and total isolation~\cite{Sirunyan:2020ycc}.
The photon reconstruction and selection efficiency for the chosen ``medium'' working point in simulation is on average 80\%. 
The electron, muon, and photon reconstruction efficiencies are corrected as a function of the \pt and $\eta$ of the reconstructed object to match the efficiency observed in data.

Furthermore, ``loose'' selection criteria are used to define control regions and to veto events with additional reconstructed leptons and photons.
With respect to the tight electron selection, the transverse momentum requirement is relaxed to $\pt>15\GeV$, the threshold on $\relIso(\Pe)$ to the range of 20--25\%, depending on $\pt$ and $\eta$ of the electron candidate, and two (three) missed hits in the innermost tracker layers are allowed for electrons in the barrel (endcap) region. 
The loose muon selection is based on Ref.~\cite{Chatrchyan:2012xi} with $\pt>15\GeV$ and $\relIso(\Pgm)<0.25$.
The loose photon selection is defined by the ``medium'' photon working point without the \chIso and $\sieie$ requirements~\cite{Sirunyan:2020ycc}. 

Jets are reconstructed by clustering PF candidates using the anti-\kt algorithm~\cite{Cacciari:2008gp,Cacciari:2011ma} with a distance parameter of 0.4.
Selected jets are required to satisfy $\pt>30\GeV$ and $\abseta<2.4$.
Contributions to the clustered energy from pileup interactions are corrected for by requiring charged-hadron candidates to be associated with the PV and
an offset correction for the contribution from neutral hadrons falling within the jet area is subtracted from the jet energy.
Corrections to the jet energy scale~(JES) are applied in simulation and data. 
The jet energy resolution~(JER) is corrected in simulation to match the resolution observed in data~\cite{Khachatryan:2016kdb}.

Jets originating from the hadronization of \PQb~quarks are identified (\PQb~tagged) with a deep neural network algorithm \cite{Sirunyan:2017ezt} based on tracking and secondary vertex information. 
A working point is chosen such that the efficiency to identify
the \PQb~jet is 55--70\% for a jet \pt of 20--400\GeV. 
The misidentification rate in this \pt range is 1--2\% for light-flavor and gluon jets, and up to 12\% for charm quark jets. 
A correction is applied to the simulation to match the \PQb~tagging efficiencies observed in data.

The missing transverse momentum vector, \ptvecmiss, is defined as the projection onto the plane perpendicular to the beams of the negative vector momentum sum of all PF candidates in an event. 
The JES and JER corrections are included in the \ptvecmiss computation. Its magnitude is referred to as \ptmiss. 

\section{Fiducial phase space definition and photon classification}\label{sec:sel_fiducial}

The fiducial region of the analysis is defined at the particle level by applying an event selection to the stable particles after the event generation, 
parton showering, and hadronization, but before the detector simulation.

Electrons~(muons) must have $\pt>35~(30)\GeV$ and $\abseta<2.4$, and must not originate from hadron decays.
To account for final-state photon radiation, the four-momenta of photons inside a cone of $\DR=0.1$ are added to the lepton before the lepton selection~\cite{Sirunyan:2018owv}.
Events with leptonically decaying $\tau$~leptons in the decay chain of the top quark are considered signal.

Photons are selected if they do not originate from hadron decays, satisfy $\ptG>20\GeV$ and $\etaG<1.4442$, and are found outside a cone of $\DR=0.4$ around the leptons.
An isolation requirement is applied by removing photons with stable particles (except neutrinos) found within a cone of $\DR=0.1$ that satisfy $\pt>5\GeV$.

Particle-level jets are clustered using the anti-\kt algorithm with a distance parameter of 0.4, using all final-state particles, excluding neutrinos. 
Jets must satisfy $\pt>30\GeV$ and $\abseta<2.4$. 
A ghost matching method~\cite{Cacciari:2007fd} is used to determine the flavor of the jets, with those matched to \PQb~hadrons tagged as \PQb~jets.
Finally, the overlap of jets and other candidates is removed by excluding jets with $\DR\leq0.4~(0.1)$ to lepton (photon) candidates.
A summary of the object definitions at particle level is provided in Table~\ref{tab:fidPS}.

The fiducial region is constructed by requiring exactly one photon, exactly one lepton, and three or more jets among which at least one must be \PQb~tagged.
The inclusive fiducial cross section, predicted with \MGvATNLO at NLO in QCD, is $773\pm 135$\fb. 
The NLO effects in the decay of the top quarks are not included in this calculation.

{\renewcommand{\arraystretch}{1.2}
\begin{table}[t]
    \topcaption{Overview of the definition of fiducial regions for various objects at particle level. A photon is isolated, if there are no stable particles (except neutrinos) with $\pt>5\GeV$ within a cone of $\DR=0.1$.
    }\label{tab:fidPS}
    \centering
    \begin{tabular}{cccc}
                 Photon & \Pe (\Pgm) & Jet & \PQb~jet \\
    \hline
          $\pt>20\GeV$ & $\pt>35~(30)\GeV$ & $\pt>30\GeV$ & $\pt>30\GeV$ \\
          $\abseta<1.4442$ & $\abseta<2.4$ & $\abseta<2.4$ & $\abseta<2.4$ \\
             no hadronic origin & no hadronic origin & $\DR(\text{jet}, \ell)>0.4$   & $\DR(\PQb~\text{jet}, \ell)>0.4$ \\
             $\dRlg>0.4$                             &                    & $\DR(\text{jet}, \gamma)>0.1$ & $\DR(\PQb~\text{jet}, \gamma)>0.1$ \\
              isolated          &                    &                    & matched to \PQb~hadrons \\
    \end{tabular}
\end{table}
}

To facilitate the estimation of backgrounds with nonprompt and misidentified photons, a photon categorization is based on the matching between the reconstructed photon and simulated particles.
Reconstructed photons are matched in \DR{} to the corresponding generator-level particle from the primary interaction.
The maximum \DR{} considered for matching is 0.3 and the \ptG is required to be within 50\% of the matched particle.
Simulated events with a reconstructed photon are subsequently classified into three categories based on the matched generator particle.
In the ``genuine photon'' category, the reconstructed photon is matched to a generated photon that originates from a lepton, a \PW boson, or a quark.
In the ``misidentified electron'' category, the photon is matched to an electron.
The ``nonprompt photon'' category is comprised of events where the photon is matched to a generated photon that originates from a hadron (71\%), 
or in absence of a match to a generated photon or electron. 
This category thus includes contributions with misidentified photons and photons that originate from pileup interactions~(29\%).

\section{Analysis strategy}\label{sec:ana_strategy}

\subsection{Signal and control region definitions}\label{sec:sel_reco}

\begin{figure}[th!b]
    \centering
    \includegraphics[width=0.31\textwidth]{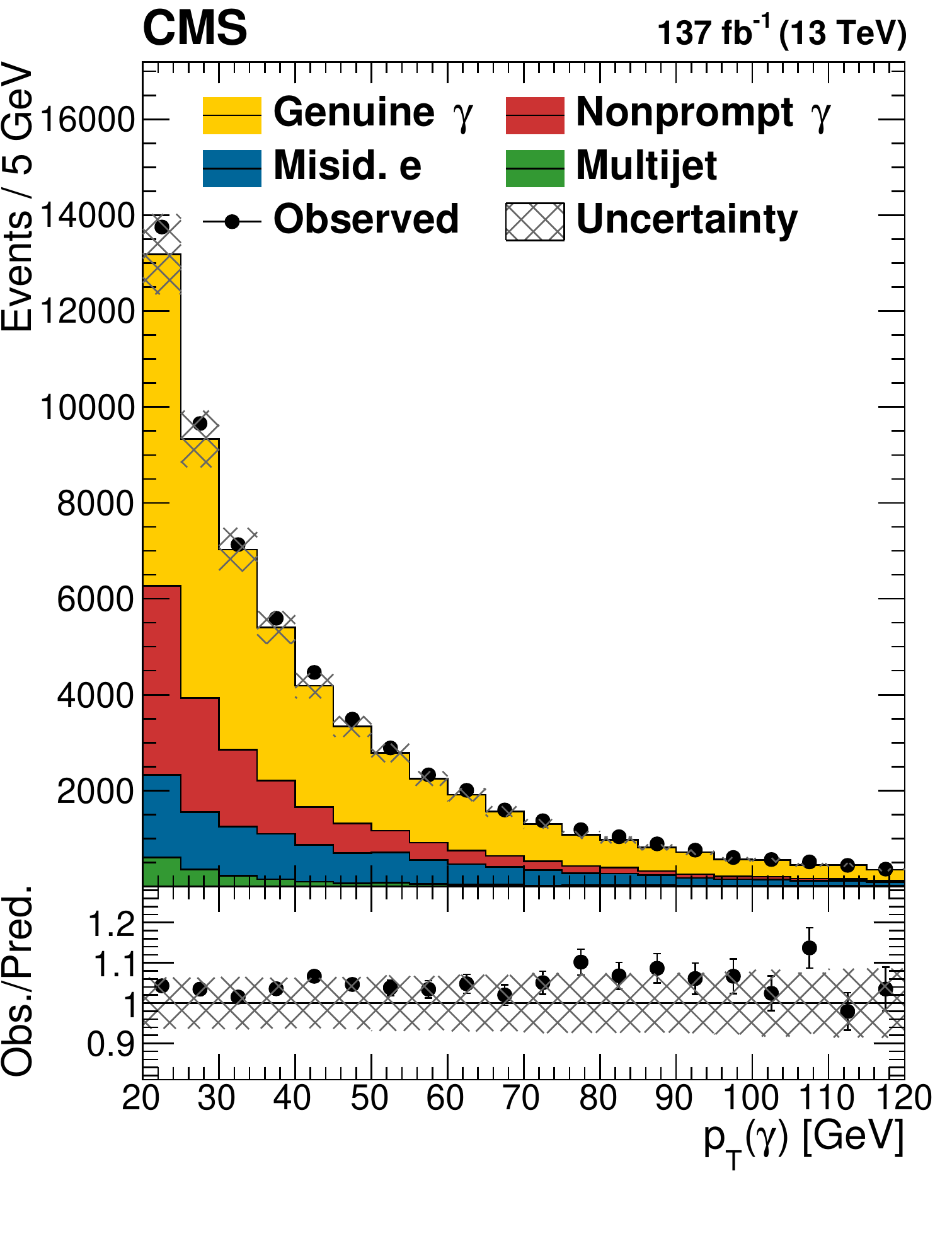}
    \includegraphics[width=0.31\textwidth]{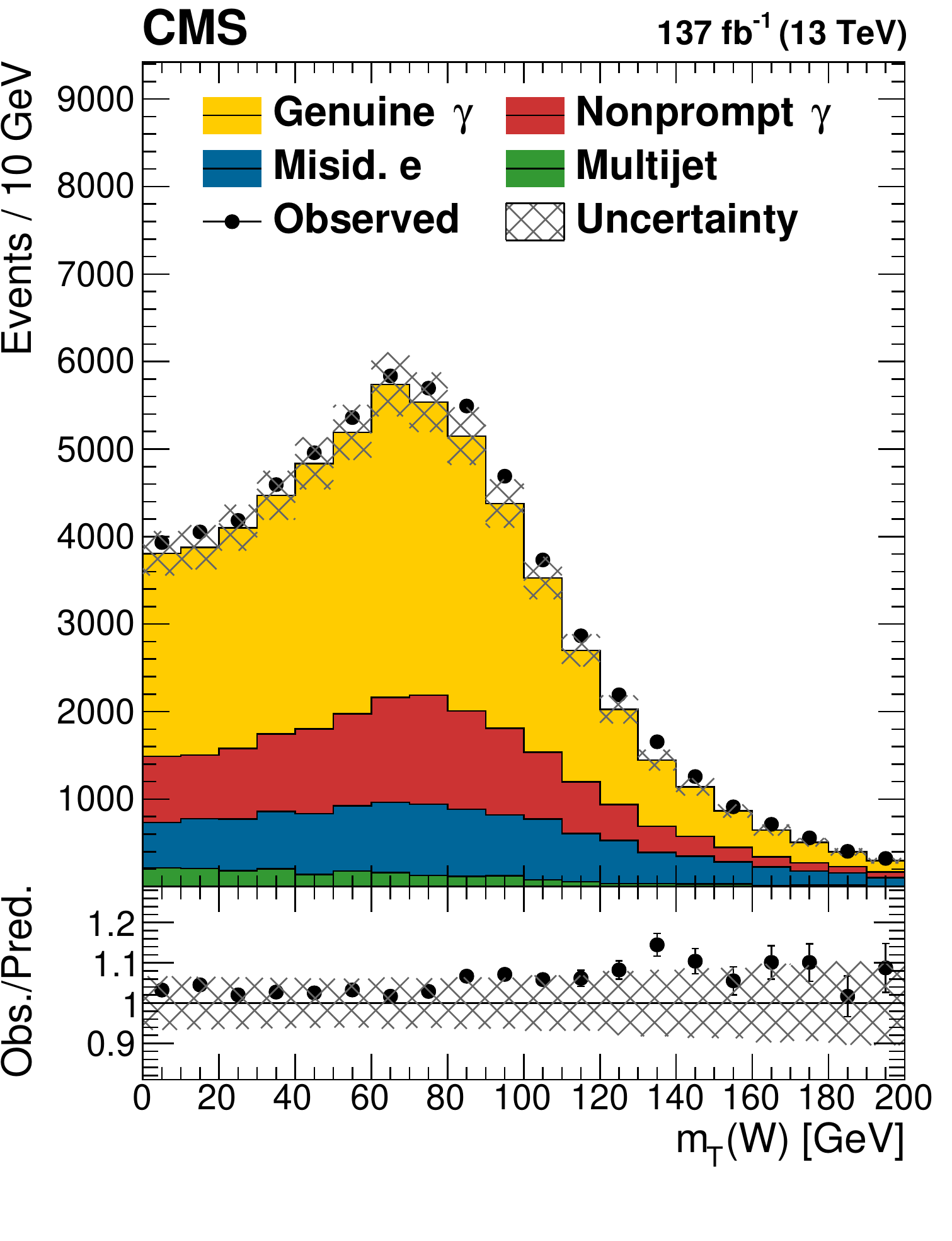}
    \includegraphics[width=0.31\textwidth]{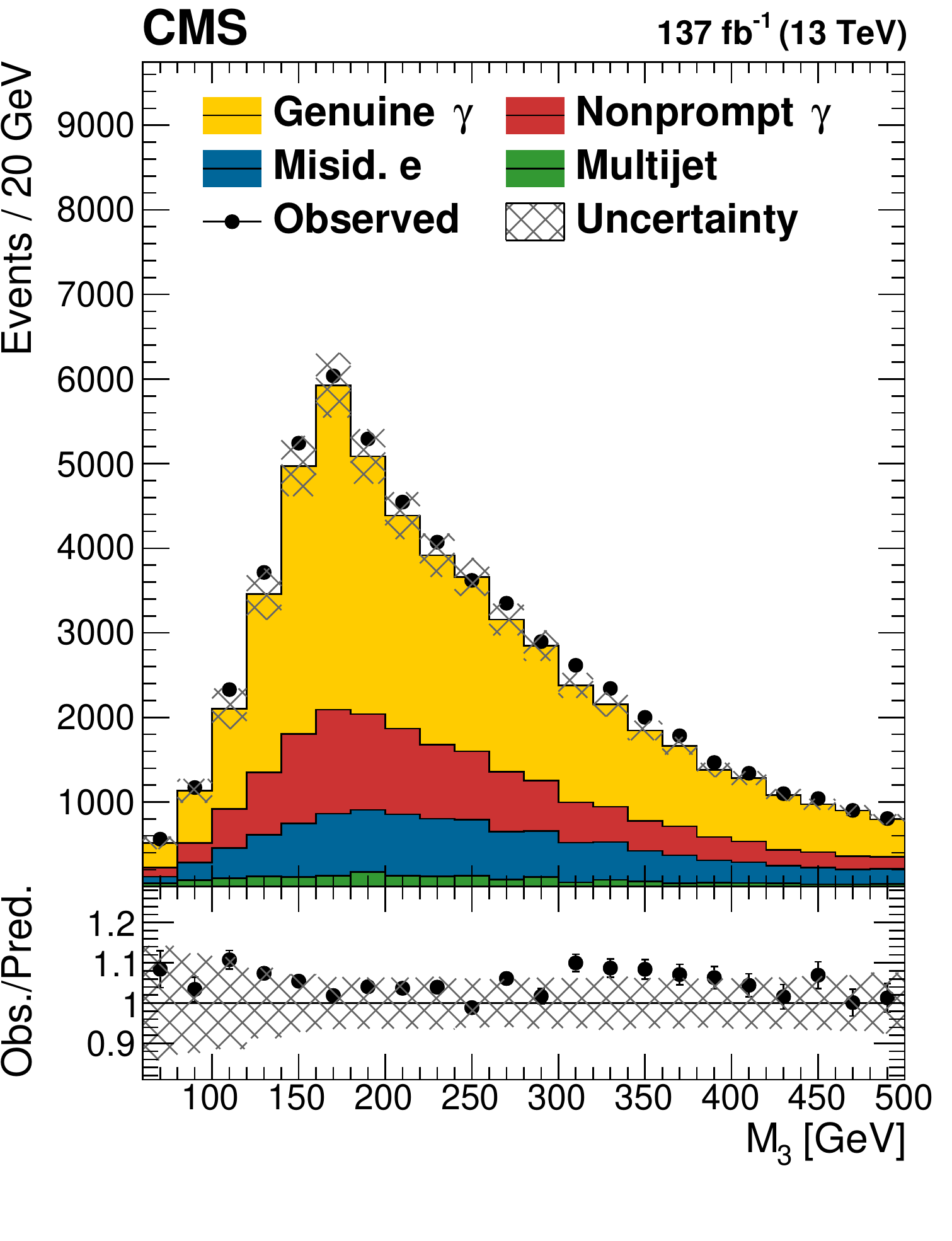}\\
    \includegraphics[width=0.31\textwidth]{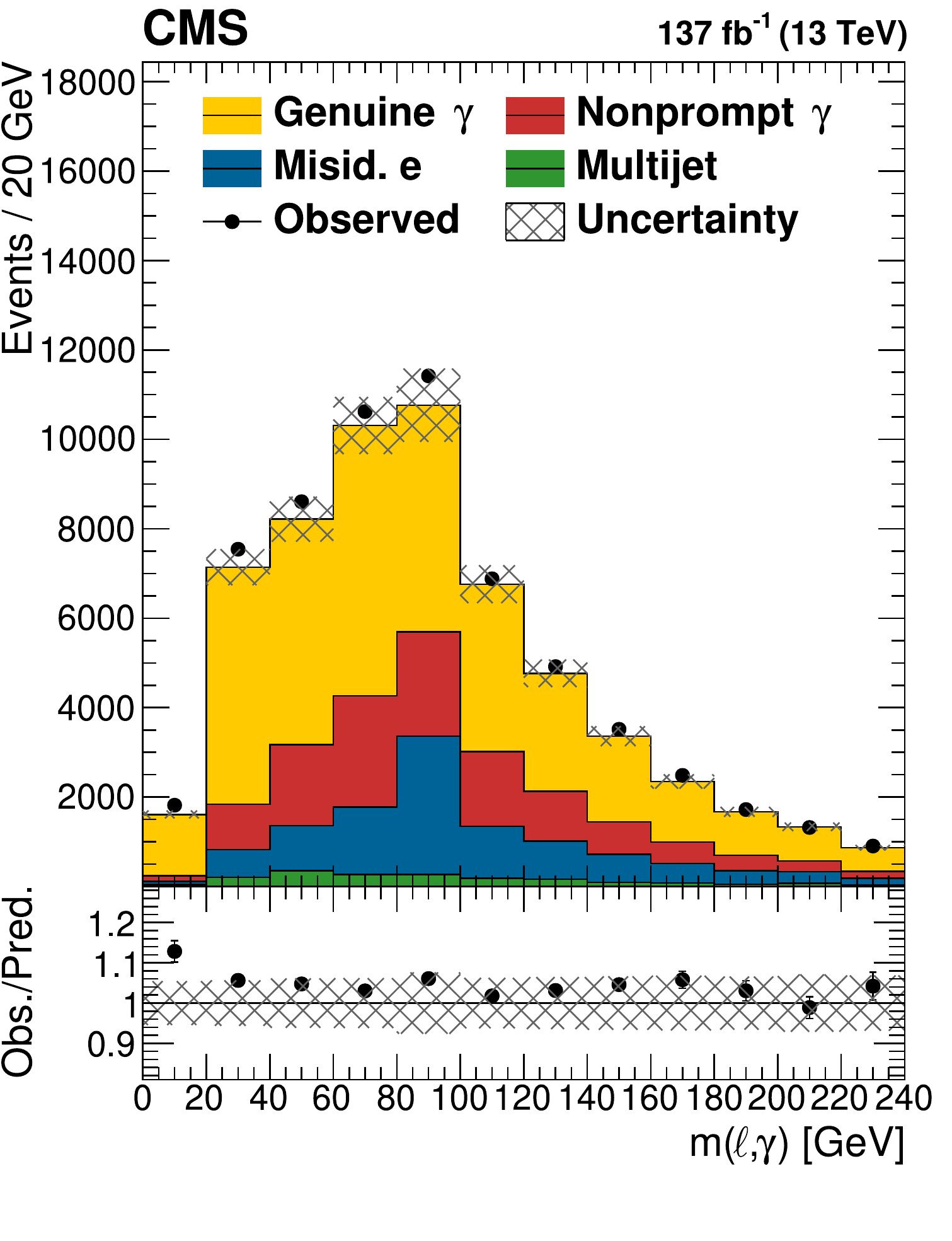}
    \includegraphics[width=0.31\textwidth]{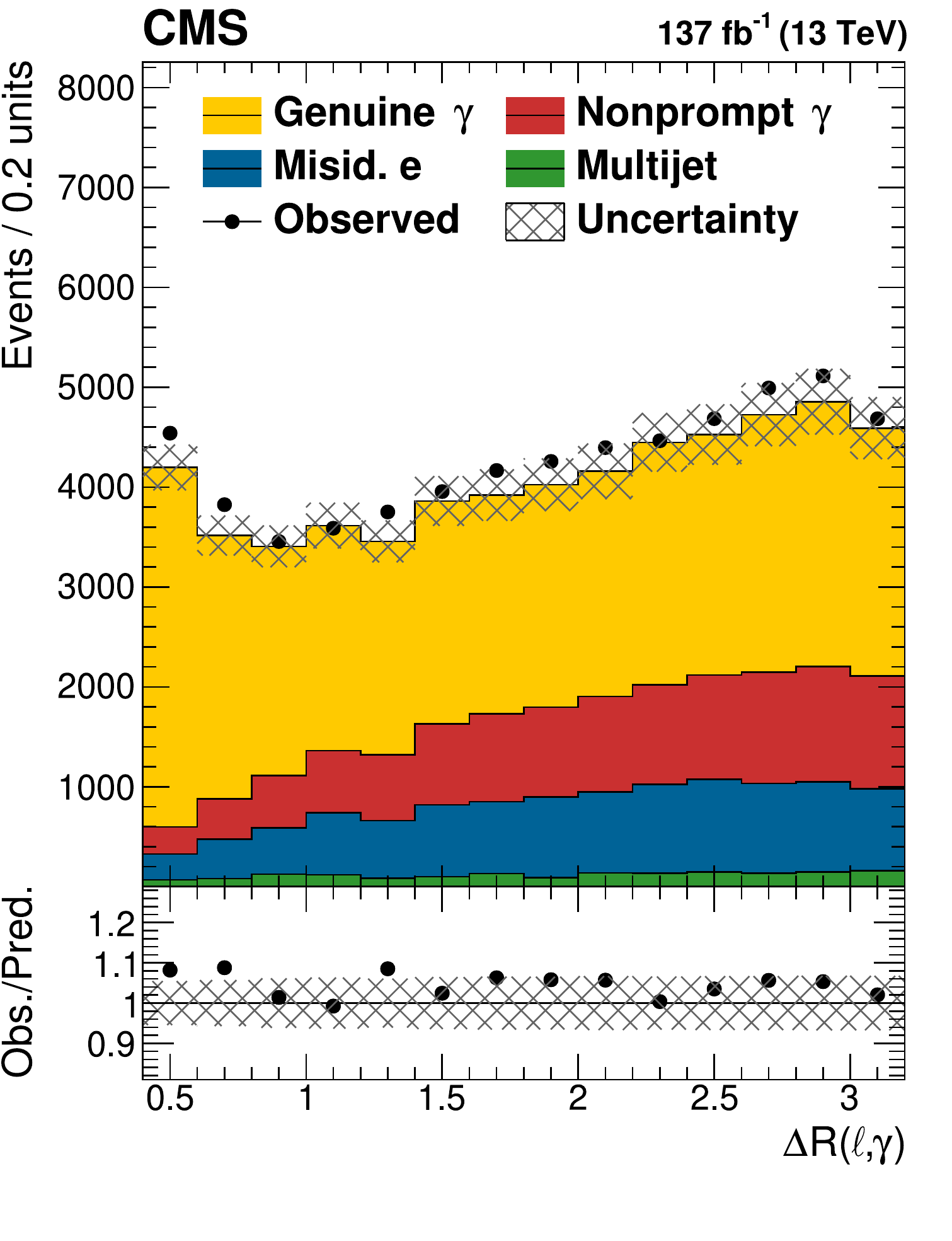}
    \includegraphics[width=0.31\textwidth]{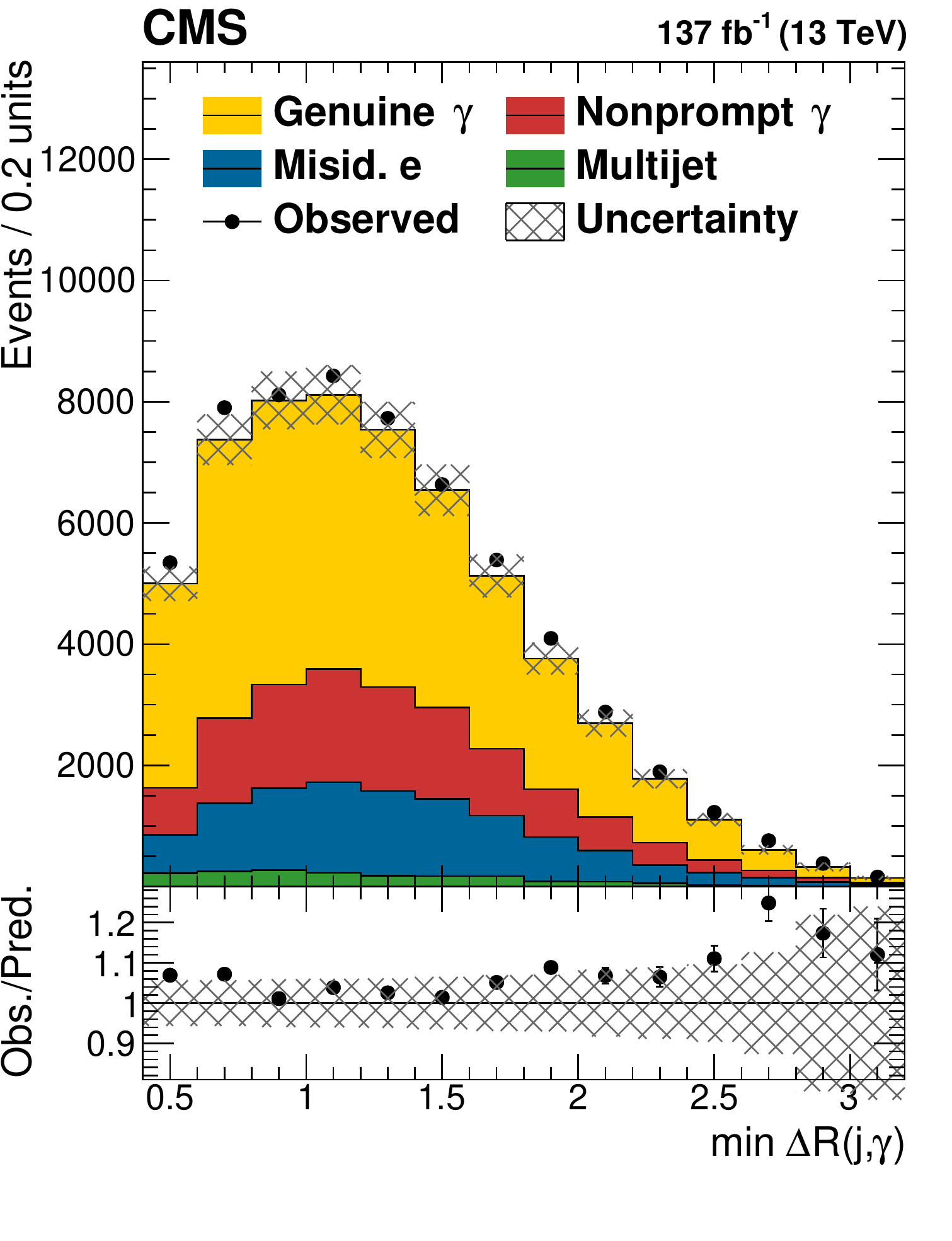}
  \caption{ Distribution of \ptG, the transverse mass \mtw of the \PW boson candidate, the three-jet invariant mass \Mthree (upper row);
            the invariant mass \mlg of the lepton and the photon, the angular separation \dRlg of the lepton and the photon, and the minimal angular separation $\textrm{min}\,\DR(j, \gamma)$ of the photon and all jets (lower row) in the SR3p region.  
The backgrounds are normalized according to the methods described in Sec.~\ref{sec:bkg} and the pre-fit systematic uncertainties are shown as a hatched band.
The lower panels show the ratio of the observed to the predicted event yields. 
}
  \label{fig:postfit}
\end{figure}

The \ttg process typically produces events with several jets, up to two \PQb-tagged jets, and an isolated photon with large \pt.
The measurement is performed in signal regions with exactly one lepton ($\nLep=1$), exactly one photon~($\nG=1$), and at least three jets ($\nJet\geq 3$), among which at least one is \PQb~tagged ($\nBTag\geq 1$).
Events with additional leptons or photons passing the loose selection are removed. 
The measurement is performed in the $\nJet=3$ and $\geq$4 signal selections, denoted by SR3 and SR4p, respectively.
Signal events with a jet failing the identification criteria thus enter the SR3 region.
The $\nJet\geq3$ selection is denoted by SR3p. 
For illustration, Fig.~\ref{fig:postfit} shows some kinematic distributions in the SR3p region where the simulated event samples are categorized according to the origin of the photon. 
The backgrounds are normalized according to the methods described in Sec.~\ref{sec:bkg} and the pre-fit systematic uncertainties are shown as a hatched band.
In this figure, \Mthree denotes the invariant mass of the three-jet combination among all identified jets that maximizes the magnitude of the vector \pt sum~\cite{Chatrchyan:2011ew}.
This choice preferentially captures the hadronic top quark decay products. 
{\renewcommand{\arraystretch}{1.10}
\begin{table}[ht!]
    \topcaption{Overview of signal and control regions. 
    }\label{tab:SRCRoverview}
    \centering
    \begin{tabular}{llccccl}
        \multicolumn{2}{c}{Region}  & \nLep & \nJet   & \nBTag  & \nG       & Other requirements\\
        \hline
\multirow{2}{*}{SR3p} & SR3          & 1     & 3 & $\geq$1       & 1   &  \\
                       & SR4p          & 1     & $\geq$4 & $\geq$1       & 1   & \\[10pt]
\multirow{4}{*}{LM3p} & \multirow{2}{*}{LM3}   & \multirow{2}{*}{1}  & \multirow{2}{*}{3} & \multirow{2}{*}{0}  & \multirow{2}{*}{1} & $m(\Pe,\gamma)<m_{\PZ}-10\GeV$, \\
        &                      &                     &                    &                     &                    & $m(\Pgm,\gamma)< m_{\PZ}$\\
        & \multirow{2}{*}{LM4p}   & \multirow{2}{*}{1}  & \multirow{2}{*}{$\geq$4} & \multirow{2}{*}{0}  & \multirow{2}{*}{1} & $m(\Pe,\gamma)<m_{\PZ}-10\GeV$, \\
        &                      &                     &                    &                     &                    & $m(\Pgm,\gamma)< m_{\PZ}$\\[10pt]
\multirow{4}{*}{HM3p} & \multirow{2}{*}{HM3}   & \multirow{2}{*}{1}  & \multirow{2}{*}{3} & \multirow{2}{*}{0}  & \multirow{2}{*}{1} & $m(\Pe,\gamma)>m_{\PZ}+10\GeV$, \\
        &                      &                     &                    &                     &                    & $m(\Pgm,\gamma)>m_{\PZ}$\\
        & \multirow{2}{*}{HM4p}   & \multirow{2}{*}{1}  & \multirow{2}{*}{$\geq$4} & \multirow{2}{*}{0}  & \multirow{2}{*}{1} & $m(\Pe,\gamma)>m_{\PZ}+10\GeV$, \\
        &                      &                     &                    &                     &                    & $m(\Pgm,\gamma)>m_{\PZ}$\\[10pt]
\multirow{2}{*}{misDY3p} & misDY3          & 1     & 3 & 0       & 1   & $\abs{m(\Pe,\gamma)-m_{\PZ}}\leq10\GeV$ \\
                       & misDY4p          & 1     & $\geq$4 & 0       & 1   & $\abs{m(\Pe,\gamma)-m_{\PZ}}\leq10\GeV$\\
    \end{tabular}
\end{table}
}

The data-based estimation procedures for the dominant background sources are described in Sec.~\ref{sec:bkg}.
The simulation predicts a significant background contribution from nonprompt photons (23\%), misidentified electrons (19\%), and a small contribution from multijet events in the SR3p region. 
The nonprompt photon contribution is estimated using background-enriched control regions with relaxed criteria on \chIso and $\sieie$.
The multijet contributions to the signal and control regions are estimated by rescaling suitable normalized distributions (templates) obtained from background-enriched high-$\relIso(\ell)$ sidebands.
The misidentified electron background is estimated in a $\nBTag=0$ region where the invariant mass of the electron and photon candidates~($m(\Pe,\gamma)$) is consistent with the \PZ boson hypothesis~\cite{Zyla:2020zbs} within 10\GeV, \ie, $\abs{m(\Pe,\gamma)-m_{\PZ}}\leq 10\GeV$, where $m_{\PZ}$ is the \PZ boson mass.  
The control region is denoted by misDY3~(misDY4p) for $\nJet=3$~($\geq$4). 
The \WGamma and the \ZGamma processes contribute events with genuine photons to both the signal regions and the misDY3 and misDY4p control regions. 
In the electron channel, their contribution is constrained in ``low mass''~(LM) and ``high mass''~(HM) regions, defined by $m(\Pe,\gamma)<m_{\PZ}-10\GeV$ and $m(\Pe,\gamma)>m_{\PZ}+10\GeV$, respectively.
In the muon channel, the LM~(HM) region is defined by $m(\Pgm,\gamma) < m_{\PZ}$~($m(\Pgm,\gamma)> m_{\PZ}$), where $m(\Pgm,\gamma)$ denotes the invariant mass of the muon and the photon.
Table~\ref{tab:SRCRoverview} provides a summary of the kinematic requirements in the signal and control regions.

\subsection{ Statistical treatment }\label{sec:stat}

The signal cross section is extracted from signal and control regions using the statistical procedure detailed in Refs.~\cite{ATL-PHYS-PUB-2011-011, Cowan:2010js}.
The observed yields, signal and background estimates in each analysis category, and
the systematic uncertainties are used to construct a binned likelihood function $L(r, \theta)$
as the product of Poisson probabilities of all bins.
The nuisances related to the systematic uncertainties in the experiment and in the modeling of signal and background processes are described by log-normal probability density functions.
The parameter $r$ is the signal strength modifier, \ie, the ratio
between the measured cross section and an arbitrary reference value of 773\fb, chosen as the nominal prediction for the inclusive fiducial cross section.
The symbol $\theta$ represents the set of nuisance parameters describing the systematic uncertainties.
The number of reconstructed \ttg signal events generated outside the fiducial phase space is scaled with the same value of $r$, \ie, no independent production cross section is assumed for this part of the signal. 

The used test statistic is the profile likelihood ratio, $q(r)=-2\ln L(r,\hat\theta_{\textrm{r}})/L(\hat{r}, \hat{\theta})$, where $\hat\theta_{\textrm{r}}$ reflects the values of the nuisance parameters that maximize the likelihood function for a signal strength modifier $r$.
The quantities $\hat{r}$ and $\hat{\theta}$ are the values that simultaneously maximize $L$.
A multi-dimensional fit is used to extract the observed cross section of the signal process, the nuisance parameters, and the uncertainties in the nuisance parameters~\cite{ATL-PHYS-PUB-2011-011, Cowan:2010js}.

The LM3, LM4p, HM3, HM4p, misDY3 and misDY4p control regions enter the likelihood fit separately for each data-taking period and lepton flavor.
In order to extract the \ptG dependence of the background with misidentified electrons, the misDY3 and misDY4p control regions are split into 7 bins separated by the \ptG thresholds 20, 35, 50, 65, 80, 120, and 160\GeV. 
The LM3, LM4p, HM3, and HM4p regions are similarly separated into 3 bins defined by the \ptG thresholds 20, 65, and 160\GeV.
The binning is chosen to obtain a statistical uncertainty in simulated background yields of less than 15\%.

The likelihood fit is performed for the inclusive cross section measurements, and separately for the differential measurement.
For the extraction of the inclusive cross section, the SR3 and SR4p signal region are divided in three \Mthree bins in the ranges 0--280, 280--420, and $>$420\GeV.
The binning in \ptG, \etaG, and \dRlg in the SR3 and SR4p signal regions for the differential measurements is provided in Section~\ref{sec:results-diff}.
The estimation of contributions from various background processes is performed using control regions binned in \ptG, which are used in all inclusive and differential measurements.

\section{Background estimation}\label{sec:bkg}
\subsection{Multijet background}\label{sec:qcd}

The probability for a multijet event to mimic the final state of the signal process is small and subject to large uncertainties. 
Therefore, the background from multijet events, comprising events with misidentified and nonprompt leptons, is estimated with a data-based procedure in sideband regions with loosened isolation criteria. 
For each \nJet requirement, a sideband region is defined by $\nBTag=0$ and requiring the lepton to pass the isolation criterion of the loose lepton working point and to fail the tight lepton selection. 
The $\nG=1$ requirement is kept.
The resulting selection is dominated by multijet events. 
After electroweak backgrounds and backgrounds with top quarks~(\WJets, $\PQt/\ttbar$, and Drell--Yan) are subtracted based on the expectation from simulation, templates for the distributions of kinematic observables are extracted.

\begin{figure}[b]
    \centering
    \includegraphics[width=0.39\textwidth]{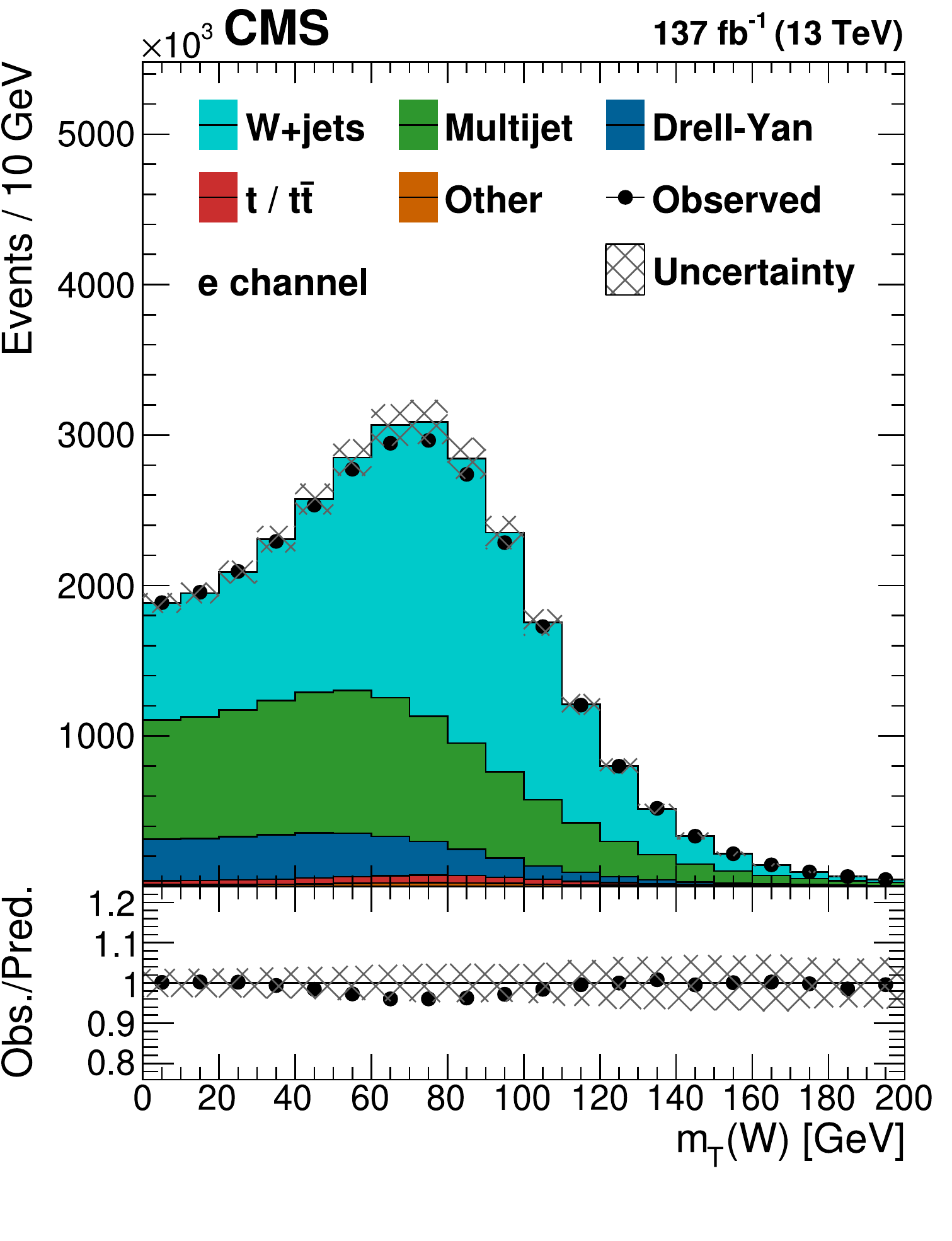}
    \hfil
    \includegraphics[width=0.39\textwidth]{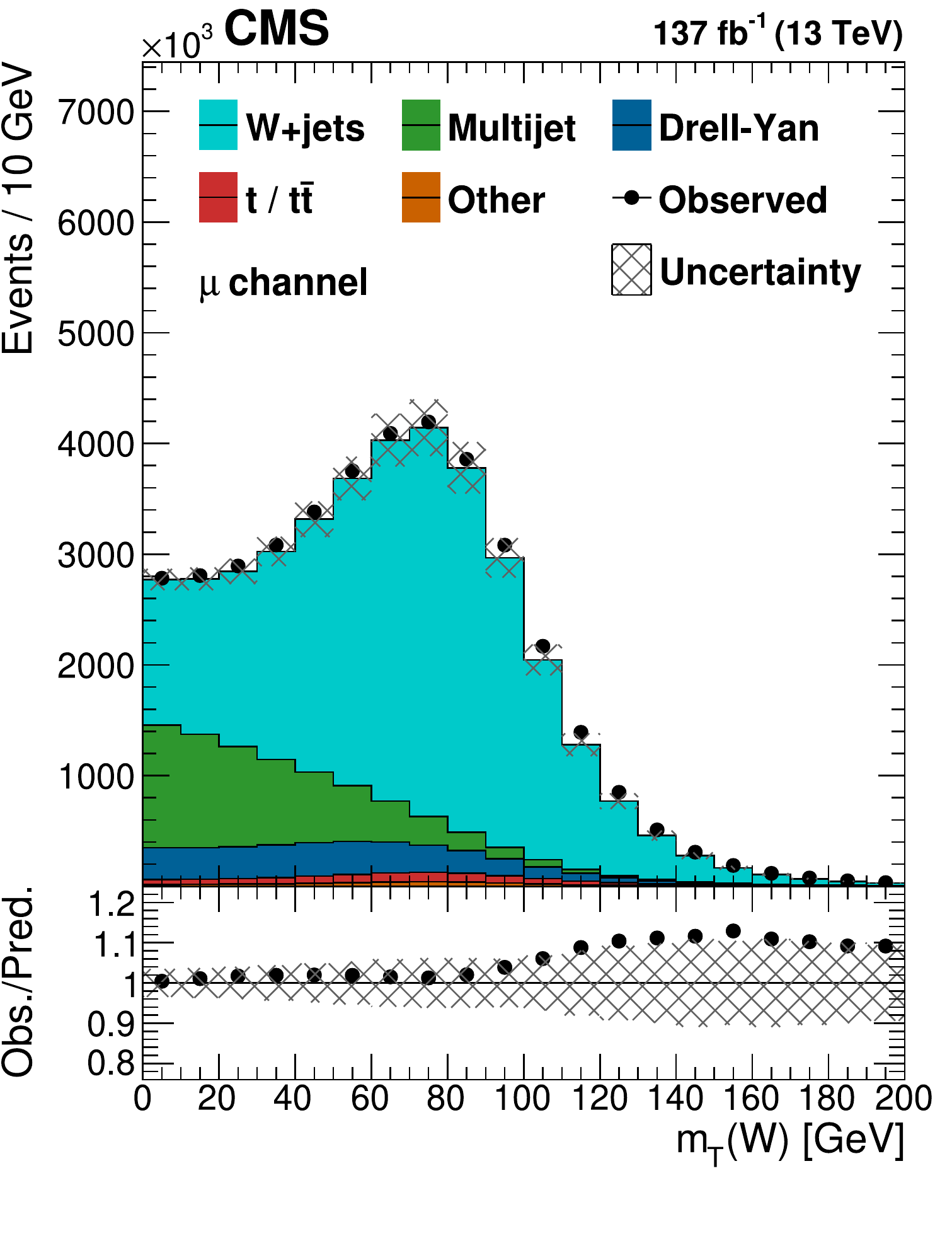}
  \caption{
    Fit result of the \mtw multijet distribution in the selection with $\nJet=2$,  $\nBTag=0$, and tightly isolated electrons~(left) and muons~(right).
    The template obtained from the selection with loosely isolated leptons~(green) and the total normalization of the electroweak and top quark background are floating in the fit. 
The lower panels show the ratio of the observed to the predicted event yields.
The pre-fit systematic uncertainties are shown as a hatched band.
}\label{fig:qcd}
\end{figure}

The template normalization is evaluated from a transfer factor (``TF''), defined as the ratio of the multijet event yield with tightly isolated lepton candidates to the yield with loosely isolated lepton candidates. 
It is obtained in a selection with $\nJet=2$ and $\nG=0$ by fitting the distribution of the transverse mass of the \PW~boson candidate, calculated from the formula
\begin{linenomath}
\begin{equation}
\mtw=\sqrt{2\pt^\ell\ptmiss[1-\cos(\Delta\phi_{\ell,\ptvecmiss})]}
\end{equation}
\end{linenomath}
where $\ell$ indicates the lepton considered in the event.
The distribution is taken from data in a $\nBTag=0$ selection with loosely isolated leptons, and electroweak and top quark backgrounds are subtracted. 
The fit is then performed in the selection with tightly isolated leptons where the total normalization of the electroweak and top quark background is left floating, while its shape is again taken from simulation.
For illustration, the fit result for the $\nJet=2$ and $\nBTag=0$ region, including the \mtw multijet distribution from the selection with loosely isolated leptons, is shown in Fig.~\ref{fig:qcd}. 

Because the efficiency of the tight lepton selection in multijet events depends on \pt and $\eta$ of the lepton,
the estimation procedure, including the TF fit, is performed in a total of 24 bins defined in these observables.
Depending on \pt and $\eta$ of the lepton, the TFs vary in the range of 0.9--3.1 (0.1--0.3) for the \Pe channel and 2.0--3.7 (0.6--1.0) for the \Pgm channel, for $\nBTag=0$ ($\geq$1).
A correction based on simulated multijet events accounts for the TF dependence on \nJet.
Finally, the multijet estimate is obtained by scaling the $\nBTag=0$ sideband templates with the corresponding TFs and accumulating the resulting predictions in the 24 bins in lepton \pt and $\eta$.
The total multijet yield is estimated at 12~(8)\% in the \Pe~(\Pgm) channel in the LM3p, HM3p and misDY3p control regions and below 0.5\% in the signal regions. 

\subsection{Nonprompt photon background}\label{sec:nonprompt}

The nonprompt photon background component is estimated from data by exploiting the difference between its distribution in the plane defined by the weakly correlated variables \sieie with \chIso,
and the corresponding distribution for genuine photons.
In a sideband with a requirement of $\sieie\geq0.011$ on the photon candidate, the expected yields with genuine photons, misidentified electrons, and multijet events are subtracted.
The sideband is used to obtain the normalization factor~$r_\textrm{SB}$, defined as the ratio of the yield passing the $\chIso<1.141\GeV$ requirement to the event yield failing it.
The estimation is obtained by multiplying $r_\textrm{SB}$ with the yield in the normalization region, defined by the nominal \sieie requirement and the inverted criteria on the photon charged hadron isolation, $\chIso>1.141\GeV$. 
The expected yields with genuine photons, misidentified electrons, and multijet events are subtracted from the observation in the normalization region.
The procedure is carried out separately for lepton flavors, \nJet selections, data-taking period, and for each bin of the differential cross section.
The deviation from unity of the double-ratio of $r_\textrm{SB}$ to the corresponding ratio in the nominal \sieie selection, stemming from the residual correlation between the two variables, is computed from simulation and it amounts to 18\%. This value is used to correct the prediction.

\subsection{Misidentified electron and genuine photon backgrounds}\label{sec:vg}

The background from electrons that are misidentified as photons is obtained from control regions defined by the requirements of $\abs{m(\Pe,\gamma)-m_\PZ}\leq 10\GeV$, and exactly three (misDY3), or four or more (misDY4p) jets.
In the simulation, these event samples have a combined purity of~58\% of Drell--Yan events with $\PZ\to\Pe\Pe$, where one of the electrons passes the photon selection criteria.
The simulated yield of the background component with a misidentified electron is multiplied by the scale factor~(SF) defined below, separately for each of the three data-taking periods.

\begin{figure}[pth]
    \centering
    \includegraphics[width=0.39\textwidth]{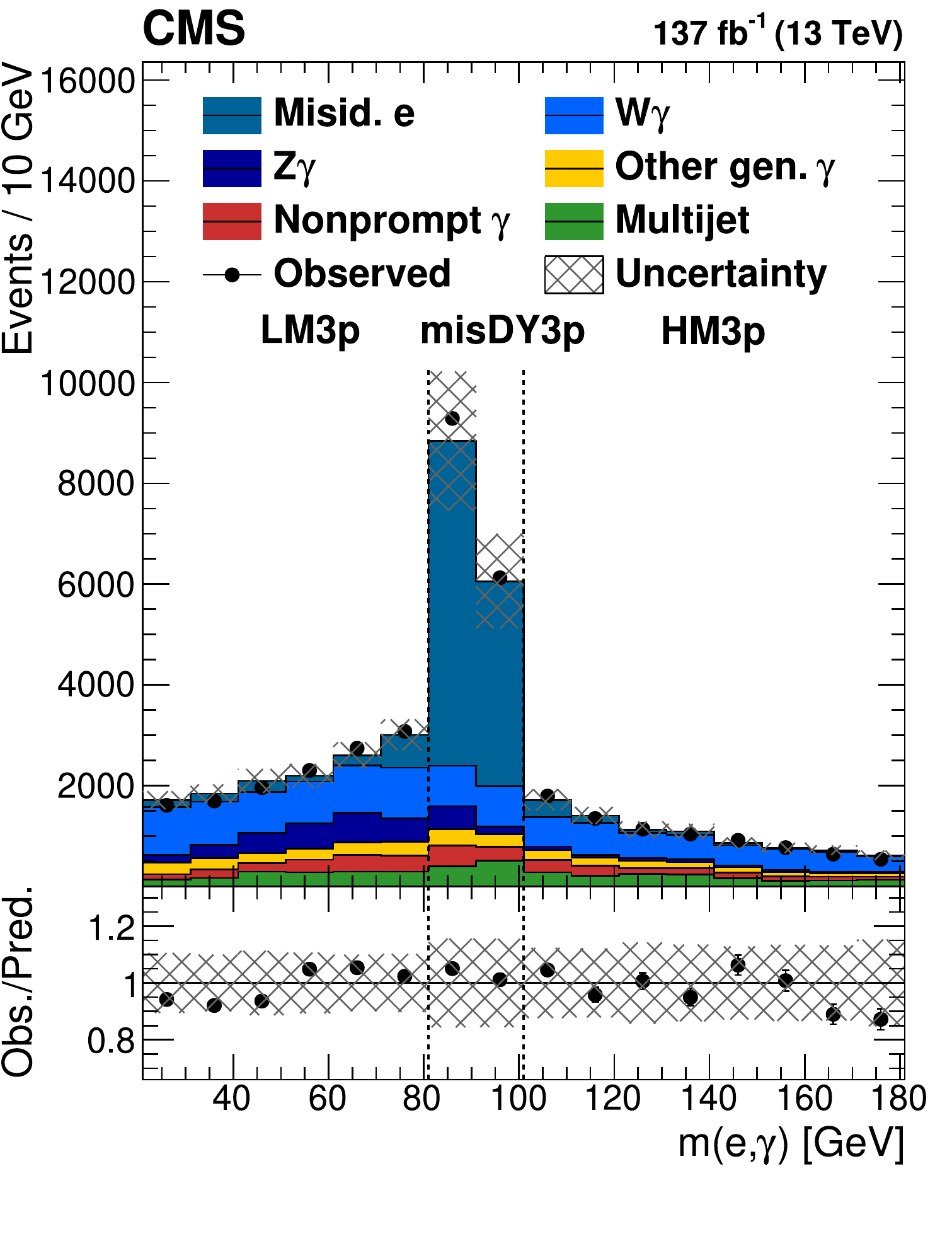}
    \hfil
    \includegraphics[width=0.39\textwidth]{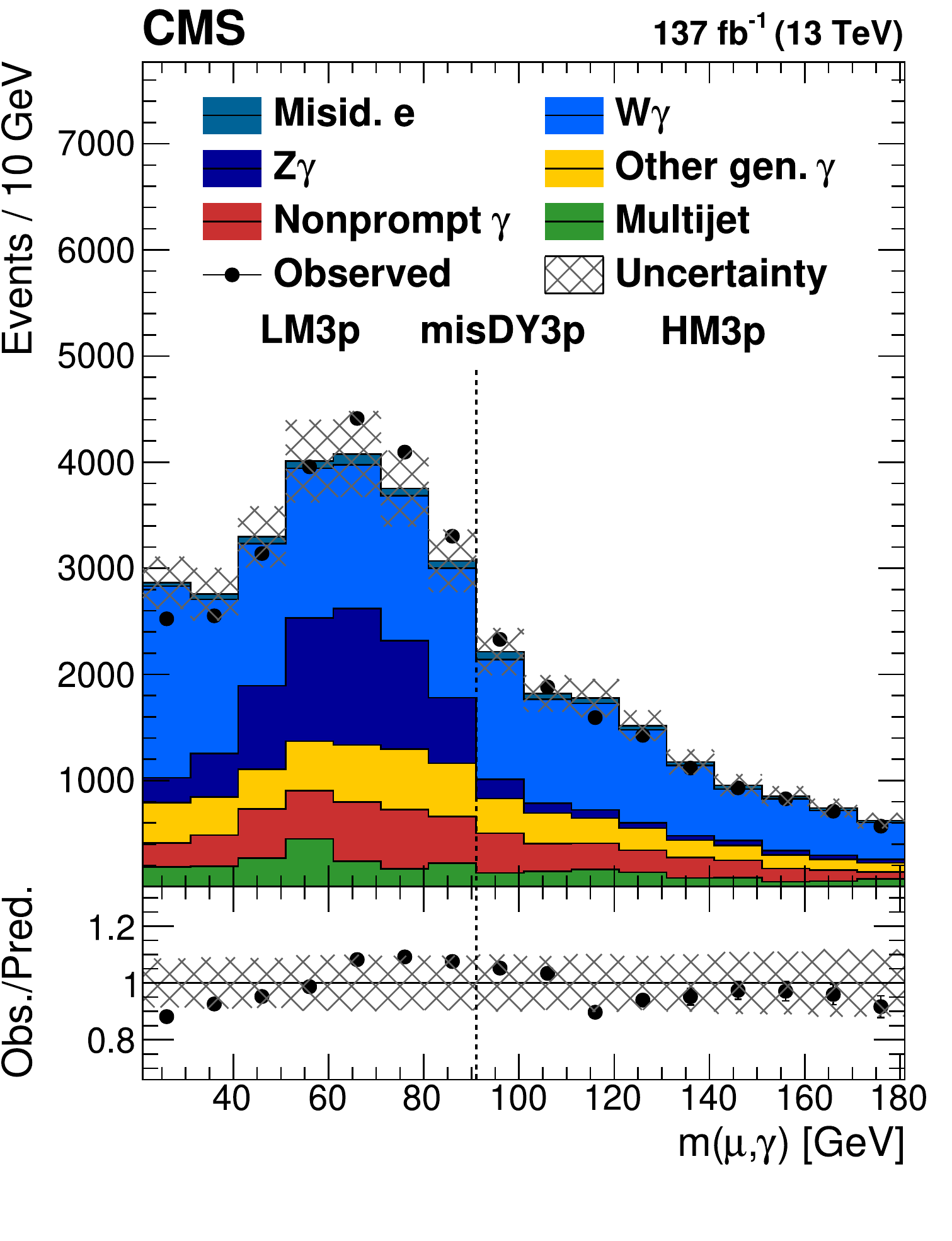}
  \caption{ Distribution of the invariant mass of the lepton and the photon, \mlg, in the ${\nJet\geq3}$, ${\nBTag=0}$ selection for the \Pe channel~(left) and the \Pgm channel~(right). 
            The genuine photon contributions of \WGamma and \ZGamma are visualized separately. 
The lower panels show the ratio of the observed to the predicted event yields.
The pre-fit systematic uncertainties are shown as a hatched band.
}
  \label{fig:VGRunIISF_fit}
\end{figure}

The \WGamma~(\ZGamma) process contributes to the LM3p regions for both lepton flavors and has a purity of 41\%~(21\%).
In the HM3p regions, the \WGamma background is dominant with a purity of~51\%. 
The SFs for the misidentified electron background and the normalization of the \WGamma and \ZGamma processes are obtained from the likelihood fit as described in Section~\ref{sec:stat}.
The fit includes the data-based multijet estimates.
The normalization of the \WGamma process is left floating and the normalization of the \ZGamma process is allowed to vary within its uncertainty. 
The resulting \mlg distributions are shown in Fig.~\ref{fig:VGRunIISF_fit} in the $\nJet\geq3$ control regions.
The background with misidentified electrons is dominant in the misDY3 and misDY4p regions close to the $m_\PZ$ peak.
A correction of 15\% to the normalization of the Drell--Yan process is measured in a data sample with two well-identified leptons satisfying $\abs{m(\ell,\ell)-m_\PZ}\leq 10\GeV$ and $\nJet\geq3$, and is included in these results. 

A summary of the extracted SF values for the misidentified electron background and the normalization of the \ZGamma and \WGamma processes, obtained from a profile likelihood fit excluding the signal regions, is provided in Table~\ref{tab:SF_RunII}. 
The observed differences in the SFs for misidentified electrons are a result of the pixel detector replacement in 2017 and its operating conditions in the three data-taking periods.  
The stability of the procedure to estimate the yields of misidentified electrons and genuine photons is assessed by repeating the fit on individual data-taking periods and separately for the $\nJet=3$ and $\geq$4 selections.
The extracted SFs from these checks agree within the uncertainties. 
For the measurements of the inclusive and differential cross sections, as well as for setting EFT limits, the SFs are determined in situ by performing the fit simultaneously with the signal regions. 
{\renewcommand{\arraystretch}{1.3}
\begin{table}[pth]
\topcaption{Extracted SFs and the total uncertainty obtained from the likelihood fit for the contribution from misidentified electrons for the three data-taking periods, and for the normalization of the \ZGamma and \WGamma background components.}\label{tab:SF_RunII}
\centering
\begin{tabular}{ll}
Scale factor & Value\\\hline
Misidentified electrons (2016) & 2.25 $\pm$ 0.29 \\
Misidentified electrons (2017) & 2.00 $\pm$ 0.27 \\
Misidentified electrons (2018) & 1.52 $\pm$ 0.17 \\
\ZGamma normalization &            1.01 $\pm$ 0.10 \\
\WGamma normalization &            1.13 $\pm$ 0.08 \\
\end{tabular}
\end{table}
}

\section{Systematic uncertainties}\label{sec:systematics}

The systematic uncertainties affecting the signal selection efficiency and background yields are summarized in Table~\ref{tab:systematics}.
The table shows the range of variations in the different bins of the analysis caused by each systematic uncertainty in the signal and background yields, as well as an estimate of the impact of each uncertainty in the measured inclusive cross section.
The table also indicates whether the uncertainties are treated as uncorrelated or fully correlated among the data-taking periods.

{\renewcommand{\arraystretch}{1.3}
\begin{table}[p!t]
\topcaption{Breakdown of the total uncertainty in its statistical and systematic components in the different signal regions.
The first column indicates the source of the uncertainty. 
The second column shows the correlation between the data-taking periods.
The third column shows the typical pre-fit uncertainties in the total simulated yields in the signal region.
The last column gives the corresponding systematic uncertainty in the \ttg cross section from the fit to the data.
}
\label{tab:systematics}
\centering
  \begin{tabular}{ clccc }
    & \multirow{2}{*}{Source}     & \multirow{2}{*}{Correlation} & \multicolumn{2}{c}{Uncertainty [\%]}\\
    & & & yield  &  $\sigma(\ttg)$\\ 
    \hline
    \multirow{11}{*}{\rotatebox{90}{\textbf{Experimental}}} & Integrated luminosity & partial & 2.3--2.5 & 1.8 \\
    & Pileup                                      & 100\%             & 0.5--2.0 & $<$0.5 \\
    & Trigger efficiency                          & \NA               & $<$0.5   & $<$0.5 \\
    & Electron reconstruction and identification  & 100\%             & 0.2--1.7 & $<$0.5 \\
    & Muon reconstruction and identification      & partial           & 0.5--0.7 & 0.7 \\
    & Photon reconstruction and identification    & 100\%             & 0.4--1.4 & 1.1 \\
    & $\pt(\Pe)$ and \ptG reconstruction          & 100\%             & 0.1--1.2 & $<$0.5 \\
    & JES                                         & partial           & 1.0--4.1 & 1.9 \\
    & JER                                         & \NA               & 0.4--1.6 & 0.6 \\
    & \PQb~tagging                                & 100\% (2017/2018) & 0.8--1.6 & 1.1 \\
    & L1 prefiring                                & 100\% (2016/2017) & 0.3--0.9 & $<$0.5 \\[.5cm]
    \multirow{5}{*}{\rotatebox{90}{\textbf{Theoretical}}}& Tune & 100\% & 0.1--1.9 & $<$0.5 \\
    & Color reconnection                 & 100\% & 0.4--3.6 & $<$0.5 \\
    & ISR/FSR                            & 100\% & 1.0--5.6 & 1.9 \\
    & PDF                                & 100\% & $<$0.5   & $<$0.5 \\
    & ME scales \muR, \muF               & 100\% & 0.4--4.7 & $<$0.5 \\[.5cm]
    \multirow{8}{*}{\rotatebox{90}{\textbf{Background}}} & Multijet normalization & 100\% & 1.3--6.5 & 0.9 \\
    & Nonprompt photon background        & 100\% & 1.2--2.7 & 1.8 \\
    & Misidentified \Pe                  & \NA   & 2.5--8.0 & 1.8 \\
    & \ZGamma normalization              & 100\% & 0.6--2.5 & 0.5 \\
    & \WGamma normalization              & 100\% & 1.0--3.5 & 2.3 \\
    & DY normalization                   & 100\% & 0.1--1.1 & 1.0 \\
    & \PQt/\ttbar normalization               & 100\% & 1.0--1.9 & 0.8 \\
    & \tWGamma modeling                  & 100\% & 1.6--4.4 & 1.6 \\
    & ``Other'' bkg. normalization       & 100\% & 0.3--1.0 & $<$0.5 \\[.5cm]
    & Total systematic uncertainty       &       &          & 6.0 \\
    & Statistical uncertainty            &       &          & 0.9 \\[.5cm]
    & Total                              &       &          & 6.0 \\
  \end{tabular}
\end{table}
}

The integrated luminosities of the 2016, 2017, and 2018 data-taking periods are individually known with uncertainties in the 2.3--2.5\% range~\cite{Sirunyan:2021qkt,CMS-PAS-LUM-17-004,CMS-PAS-LUM-18-002}, while the total Run~2 (2016--2018) integrated luminosity has an uncertainty of 1.8\%, the improvement in precision reflecting the (uncorrelated) time evolution of some systematic effects. 
The uncertainty in the inclusive cross section from these sources is, therefore, 1.8\%. 
Simulated events are reweighted according to the distribution of the number of interactions in each bunch crossing corresponding to a total inelastic \pp cross section of 69.2\unit{mb}~\cite{Sirunyan:2018nqx}.
The uncertainty in the total inelastic \pp cross section is 4.6\%~\cite{Aaboud:2016mmw} and affects the pileup estimate.
The uncertainty due to the pileup effect is about 2\% for the expected yields and less than 0.5\% for the inclusive cross section.

The uncertainties in the SFs used to match the simulated trigger selection efficiencies to the ones observed in data are propagated to the results.
From the ``tag-and-probe'' measurement~\cite{Chatrchyan:2012xi,Khachatryan:2015hwa}, an uncertainty of up to 0.5\% is assigned to the yields obtained in simulation.
Lepton selection efficiencies are measured in bins of lepton \pt and $\eta$, and are found to be in the range 50--80 (75--85)\% for electrons~(muons). 
These measurements are performed separately in data and simulation and their ratio is used to scale the yields obtained in the simulation.
The impact of these uncertainties on the inclusive cross section is 0.5~(0.7)\% for the electron~(muon) channel.

In the barrel section of the ECAL, an energy resolution of about 1\% is achieved for unconverted or late-converting highly energetic photons in the tens of \GeV energy range.
The energy resolution of the remaining barrel photons is about 1.3\% up to $\abs{\eta} = 1$, changing to about 2.5\% at $\abs{\eta} = 1.4$~\cite{Khachatryan:2015iwa}. 
Uncertainties in the photon energy scale and resolution are measured with electrons from \PZ boson decays, reconstructed using information exclusively from the
ECAL~\cite{Khachatryan:2015iwa,Sirunyan:2020ycc}. 
Additionally, an event sample enriched in $\Pgmp\Pgmm\gamma$ events is used to measure an SF correcting the efficiency of the electron veto~\cite{Chatrchyan:2013dga}. 
The total uncertainty in the photon energy and identification amounts to 1.1\% for the inclusive cross section, and reaches 2\% for $\ptG>100\GeV$.

The jet energy calibration accounts for the effects of pileup, the uniformity of the detector response, and residual data-simulation jet energy scale differences corrected
using Drell--Yan, dijet, and $\gamma$+jet events.
Uncertainties in the JES are estimated by shifting the jet energy corrections in simulation up and down by one standard deviation.
Depending on \pt and $\eta$, the uncertainty in JES varies in the range 2--5\%, leading to uncertainties in the predicted signal and background yields of 1.0--4.1\% and an impact on the inclusive cross section of 1.9\%.
The dominant components originate from the uncertainty in the jet-flavor composition in the Drell--Yan and dijet selections~(JES--FlavorQCD) and the absolute jet energy scale~(JES--Absolute)~\cite{Khachatryan:2016kdb}.
For the signal and background processes modeled via simulation, the uncertainty in the measurement is determined from the observed differences in the yields with and without the shift in jet energy corrections.
The same technique is used to calculate the uncertainties from the JER, which are found to be less than 1\%~\cite{Khachatryan:2016kdb}.
The \PQb~tagging efficiency in the simulation is corrected using SFs determined from data, separately for \cPqb~jets, \cPqc~jets, and $\cPqu\cPqd\cPqs\cPg$~jets~\cite{Chatrchyan:2012jua,Sirunyan:2017ezt}.
These are estimated separately for correctly and incorrectly identified jets, and each results in an uncertainty of about 0.8--1.6\% in the yields in the signal regions, depending on \nBTag.

During the 2016 and 2017 data-taking periods, a gradual shift in the timing of the inputs of the ECAL L1 trigger in the forward endcap region ($\abseta>2.4$) led to a specific inefficiency (labeled ``L1 prefiring'' in Table~\ref{tab:systematics}). 
A correction for this effect was determined using an unbiased data sample and is found to be relevant in events with jets with $2.4<\abseta<3.0$ and $\pt>100\GeV$. 
While no reconstructed objects at this $\eta$ directly enter the measurements, it can affect the \ptmiss observable. 
A systematic variation of 20\% of this correction for affected objects leads to an uncertainty of 0.3--0.9\% in the predicted yields. 

To estimate the theoretical uncertainties from missing higher-order corrections in the signal cross section calculation, the choice of \muR and \muF are varied independently up and down by a factor of 2.
The acceptance variations are taken as the systematic uncertainty in each bin and are found to be smaller than 4.7\%.
Two independent nuisance parameters are used for the uncertainty in the choice of \muR and \muF, and their impact on the inclusive cross section measurement in the profile likelihood fit is less than 0.5\%.
A test with a single nuisance parameter, associated with the envelope of the uncertainties related to the choice of \muR and \muF, leads to negligible differences.
The different sets in the NNPDF PDF~\cite{Ball:2014uwa} are used to estimate the corresponding uncertainty in the acceptance for the cross section measurement, which is less than 0.5\%.
The scale, PDF, and \alpS uncertainties in the inclusive fiducial cross section of the \ttg process, evaluated with \MGvATNLO at NLO in QCD, amount to 17.5\%. 
The limited number of available simulated events is considered by performing the fit using the Barlow--Beeston method~\cite{Barlow:1993dm}.

In the parton shower simulation, the uncertainty from the choice of \muF is estimated by varying the scale of initial- and final-state radiation~(ISR/FSR) up and down by factors of 2 and $\sqrt{2}$, respectively, as suggested in Ref.~\cite{Skands:2014pea}.
The default configuration in \PYTHIA includes a model of color reconnection based on multiple parton interactions (MPI) with early resonance decays switched off.
To estimate the uncertainty from this choice of model, the variations of the simulated yields with different color reconnection schemes within \PYTHIA are treated as systematic uncertainties: the MPI-based scheme with early
resonance decays switched on, a gluon-move scheme~\cite{Argyropoulos:2014zoa}, and a QCD-inspired scheme~\cite{Christiansen:2015yqa}.
The total uncertainty from color reconnection modeling is estimated by taking the maximum deviation from the nominal result and amounts to less than 0.5\% in the inclusive cross section.

The \tWGamma background component amounts to at most 3.3\% of the total event yield in the SR3 and SR4p signal regions and is predicted by the \tW sample, simulated with \POWHEG at NLO precision. 
To account for uncertainties in the \tWGamma modeling, we treat the difference between the nominal prediction from the parton shower in the \tW sample, normalized to NNLO, and a prediction obtained from \MGvATNLO at LO for the $2\to 3$ process as an uncertainty. 
For the SR3 (SR4p) signal regions, the differences of the total \tWGamma contribution are less than 44\% (30\%) in the \ptG bins, less than 34\% (27\%) in the \etaG bins, and less than 19\% (17\%) in the \dRlg bins and lead to an uncertainty of 1.6\% in the inclusive fiducial cross section. 

The uncertainty in the normalization of the QCD multijet component is based on the variation of the TF with \nJet for different \nBTag and amounts to 50\%.
Independent uncertainties are considered for the contributions to the $\nBTag=0$ and $\geq$1 yields. 
These have a significant impact only in the LM3, LM4p, HM3, and HM4p control regions, and lead to an uncertainty of 0.9\% in the measured inclusive cross section.

The uncertainty in the nonprompt photon prediction is based on the modeling of the \chIso distribution for different requirements on \sieie and leads to an uncertainty of 1.8\% in the inclusive cross section.
The normalization of the \WGamma process is left floating in the profile likelihood. 
To account for the uncertainty in the \nJet modeling of the \ZGamma process, we include an uncertainty of 30\% in its normalization.
In the signal region, the contribution of \WGamma and \ZGamma background events generated with additional \PQb or \PQc quarks is 30\%, and we assign an uncertainty of 20\% in its normalization.
Moreover, 40~(20)\% uncertainty is assigned to the normalization of the \ZGamma~(\WGamma and misidentified electron) background in the $\nJet\geq4$ signal and control regions.
The corresponding impact of the normalization of the \ZGamma and \WGamma contributions are 0.5 and 2.3\%, respectively. 
The component with misidentified electrons leads to an uncertainty of up to 8\% in the predicted background yields with an impact on the inclusive cross section of 1.8\%.
The 8\% uncertainty in the normalization of the Drell--Yan process, the 5\% uncertainty in the \PQt/\ttbar normalization, and the uncertainties in the normalization of other small background components lead to additional uncertainties below 1\%. 

\section{Results}\label{sec:results}

\subsection{Inclusive cross section measurement}
\begin{figure}[t]
    \centering
    \includegraphics[width=\textwidth]{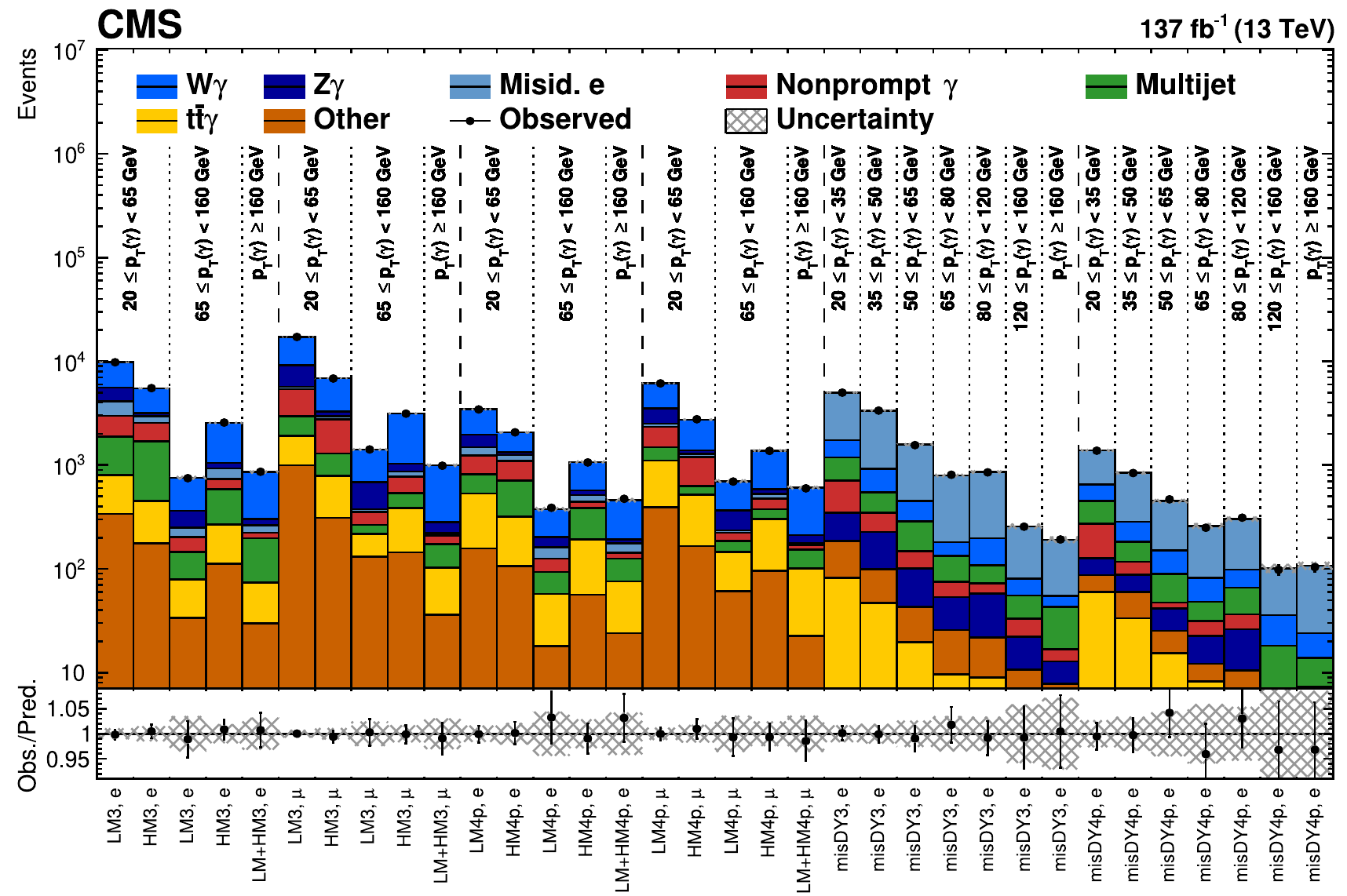}
    \caption{Fitted and observed yields in the LM3, LM4p, HM3, HM4p, misDY3, and misDY4p control regions using the post-fit values of the nuisance parameters. 
The lower panel shows the ratio of the observed to the predicted event yields.
The post-fit systematic uncertainties are shown as a hatched band.
}\label{fig:regions_RunII_CR_postfit}
\end{figure}

\begin{figure}[p]
    \centering
    \includegraphics[width=0.75\textwidth]{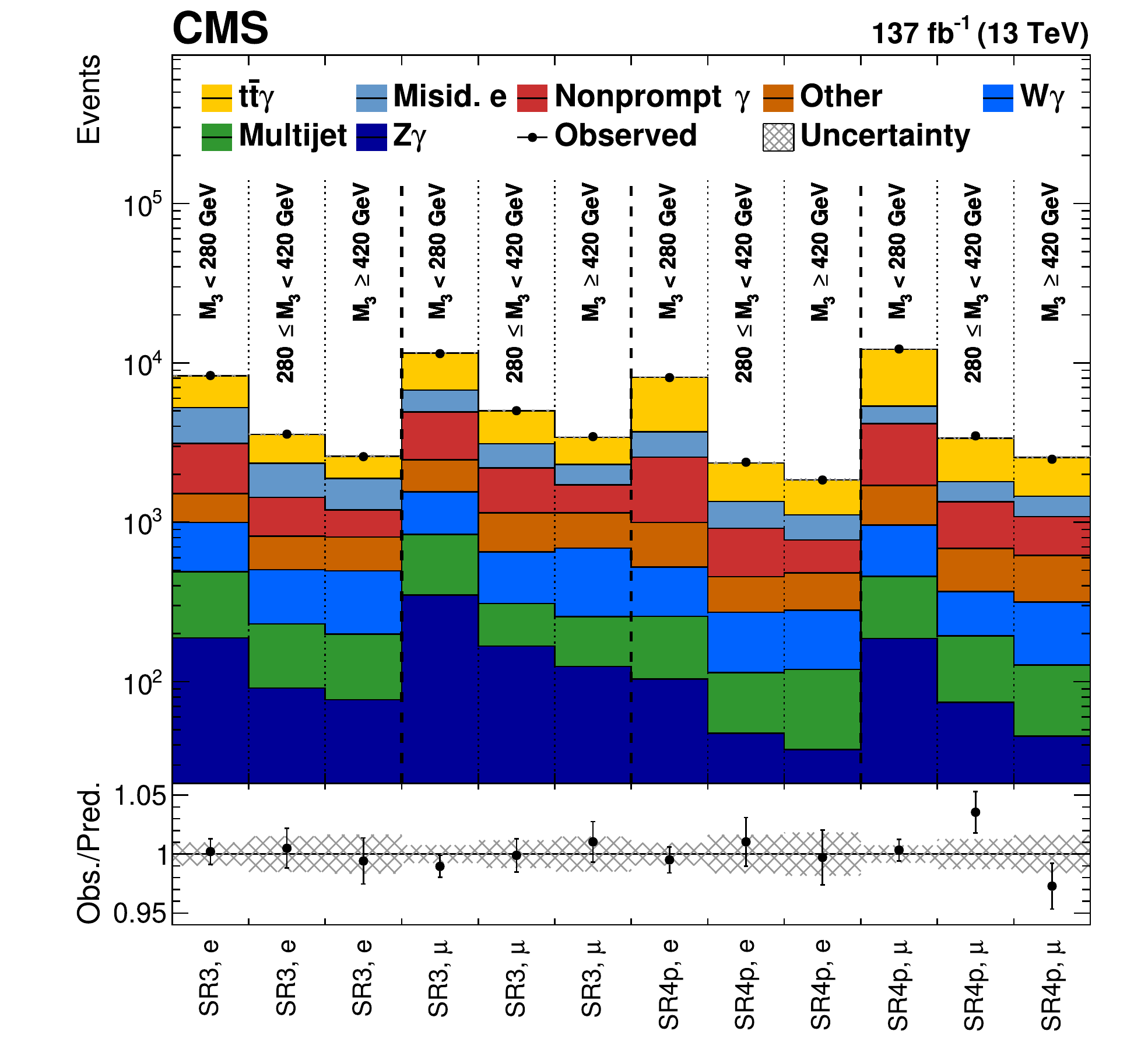}
    \caption{Fitted and observed yields in the SR3 and SR4p signal regions using the post-fit values of the nuisance parameters. 
The lower panel shows the ratio of the observed to the predicted event yields.
The post-fit systematic uncertainties are shown as a hatched band.
}\label{fig:regions_RunII_postfit_incl}
\end{figure}

\begin{table}[p]
    \setlength{\tabcolsep}{1pt}
    \caption{The observed number of events for the SR3 and SR4p signal regions in the \Pe and \Pgm channels, and the predicted yields and total post-fit uncertainties in each background component.}\label{tab:post-fit-yields}
    \centering
    \begin{tabular}{lrp{0.02\textwidth}rp{0.03\textwidth}rp{0.02\textwidth}rp{0.03\textwidth}rp{0.02\textwidth}rp{0.03\textwidth}rp{0.02\textwidth}r}
                 \multirow{2}{*}{Process} & \multicolumn{7}{c}{SR3}                              & & \multicolumn{7}{c}{SR4p} \\
                                          & \multicolumn{3}{c}{\Pe} & & \multicolumn{3}{c}{\Pgm} & & \multicolumn{3}{c}{\Pe} & & \multicolumn{3}{c}{\Pgm} \\
    \hline
    \ttg                   &    4995  & $\pm$  & 168   & &   7821  & $\pm$  & 251   & &   6174  & $\pm$  & 192   & &   9495  & $\pm$  & 280\\
    Misid. e               &    3710  & $\pm$  & 200   & &   3322  & $\pm$  & 220   & &   1904  & $\pm$  & 134   & &   2015  & $\pm$  & 153\\
    Nonprompt $\gamma$     &    2621  & $\pm$  & 107   & &   4077  & $\pm$  & 161   & &   2315  & $\pm$  & 124   & &   3580  & $\pm$  & 149\\
    Other                  &    1136  & $\pm$  & 102   & &   1866  & $\pm$  & 159   & &   857   & $\pm$  & 110   & &   1360  & $\pm$  & 166\\
    \WGamma                &    1082  & $\pm$  & 77    & &   1486  & $\pm$  & 108   & &   585   & $\pm$  & 48    & &   864   & $\pm$  & 74\\
    Multijet               &    560   & $\pm$  & 104   & &   762   & $\pm$  & 140   & &   302   & $\pm$  & 65    & &   472   & $\pm$  & 102\\
    \ZGamma                &    356   & $\pm$  & 38    & &   640   & $\pm$  & 68    & &   189   & $\pm$  & 25    & &   306   & $\pm$  & 40\\[1ex]
    Total                  &    14459 & $\pm$  & 178   & &   19976 & $\pm$  & 196   & &   12326 & $\pm$  & 150   & &   18093 & $\pm$  & 173\\[1ex]
    Observed               &  \multicolumn{3}{c}{14479}& & \multicolumn{3}{c}{19885}& &\multicolumn{3}{c}{12305} & &\multicolumn{3}{c}{18184} \\
    \end{tabular}
\end{table}

The observed data, as well as the predicted signal and background yields resulting from the likelihood fit to all signal and control regions, are shown in Figs.~\ref{fig:regions_RunII_CR_postfit} and \ref{fig:regions_RunII_postfit_incl}.
In these figures, the contributions from the three data-taking periods are summed, accounting for the correlation of the systematic uncertainties.  
The signal cross section is extracted from these categories using the statistical procedure detailed in Section~\ref{sec:stat}.
In the fit, nuisance parameters for the various systematic uncertainties and the normalization of background processes, as described in Section~\ref{sec:systematics}, are included.
The theoretical uncertainty in the inclusive fiducial cross section does not enter the likelihood fit for the inclusive or differential cross section measurements.
Using three bins in \Mthree reduces the uncertainty in the backgrounds without a hadronically decaying top quark, \eg, the misidentified electron background and the \WGamma and \ZGamma processes, and decreases the total relative uncertainty in the inclusive cross section from 6.7 to 6.0\%.
The observed number of events for the SR3 and SR4p signal regions in the \Pe and \Pgm channels, and the predicted yields and total uncertainties in each background component are listed in Table~\ref{tab:post-fit-yields}.

\begin{figure}[t]
    \centering
    \includegraphics[width=0.7\textwidth]{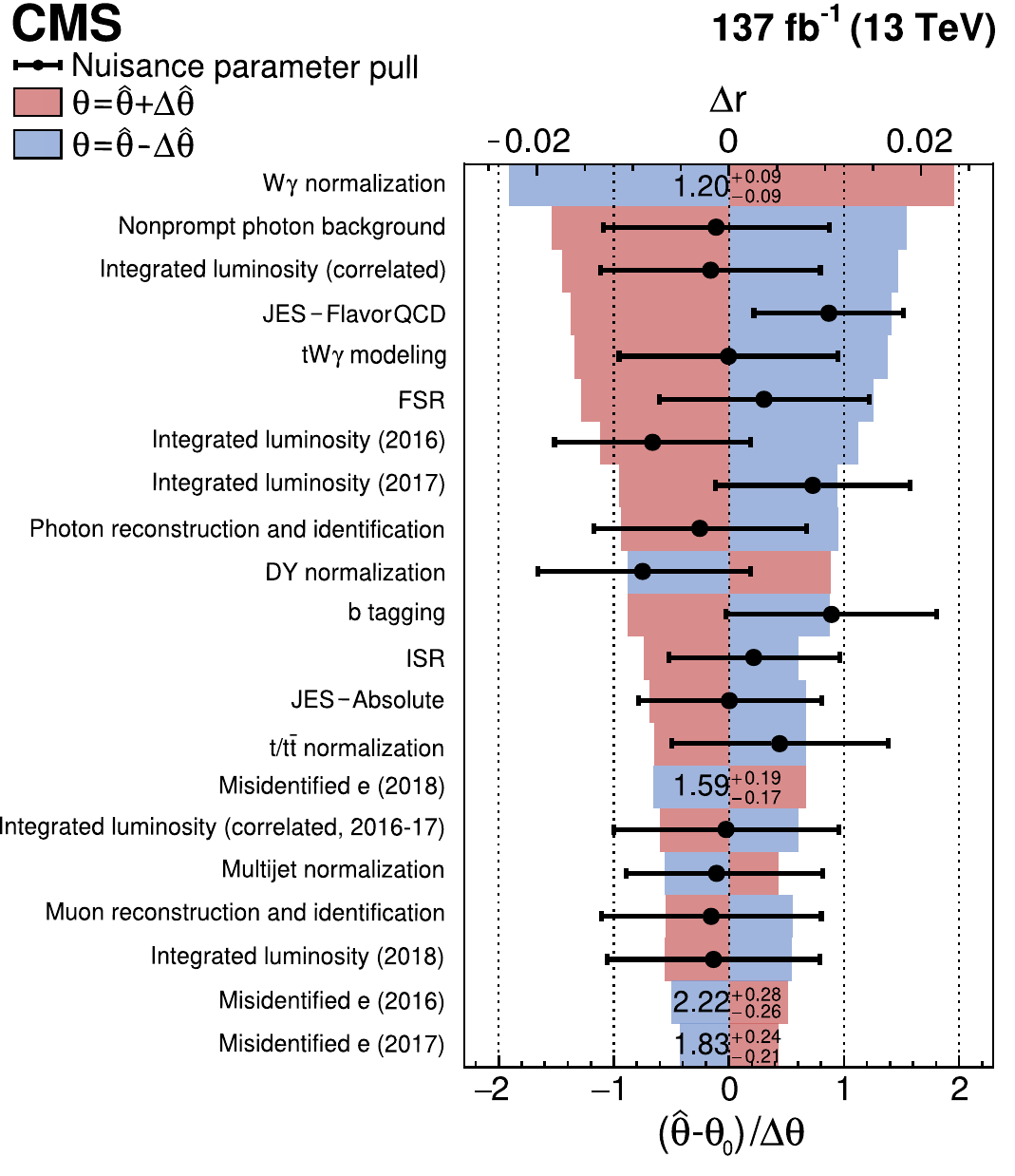}
    \caption{
Ranking of the systematic uncertainties from the profile likelihood fit used in the inclusive cross section measurement. 
For each uncertainty, the red and blue bands indicate the post-fit impact on the fit result. 
The black dots indicate the post-fit values of the nuisance parameters and the numerical values provide the extracted SFs for the misidentified electron background and the normalization of the \WGamma process.
The black lines represent the post-fit uncertainties normalized to the pre-fit uncertainties~(constraints). 
}\label{fig:pulls_incl}
\end{figure}

Figure~\ref{fig:pulls_incl} shows the ranking of the leading systematic uncertainties according to their post-fit impact on the measured inclusive cross section. 
The normalization of the \WGamma background is the largest individual contribution to the uncertainty in the inclusive cross section measurement and amounts to about 2.3\%.
The post-fit values of the nuisance parameters~(pulls) are also shown and are found to lie within the pre-fit uncertainties.
The extracted SFs and the normalizations of the \ZGamma and \WGamma backgrounds also agree with the fit result in Table~\ref{tab:SF_RunII}, obtained exclusively from the control regions.
Besides the extraction of the nuisance parameters related to the normalization of the misidentified electron component and the \ZGamma and \WGamma backgrounds, the only mild constraints are 35\% for the JES--FlavorQCD nuisance and 25\% for the scale of ISR.
They reflect the improvement of the uncertainty in the inclusive cross section induced by the binning in \Mthree.  

\begin{figure}[t]
    \centering
    \includegraphics[width=0.8\textwidth]{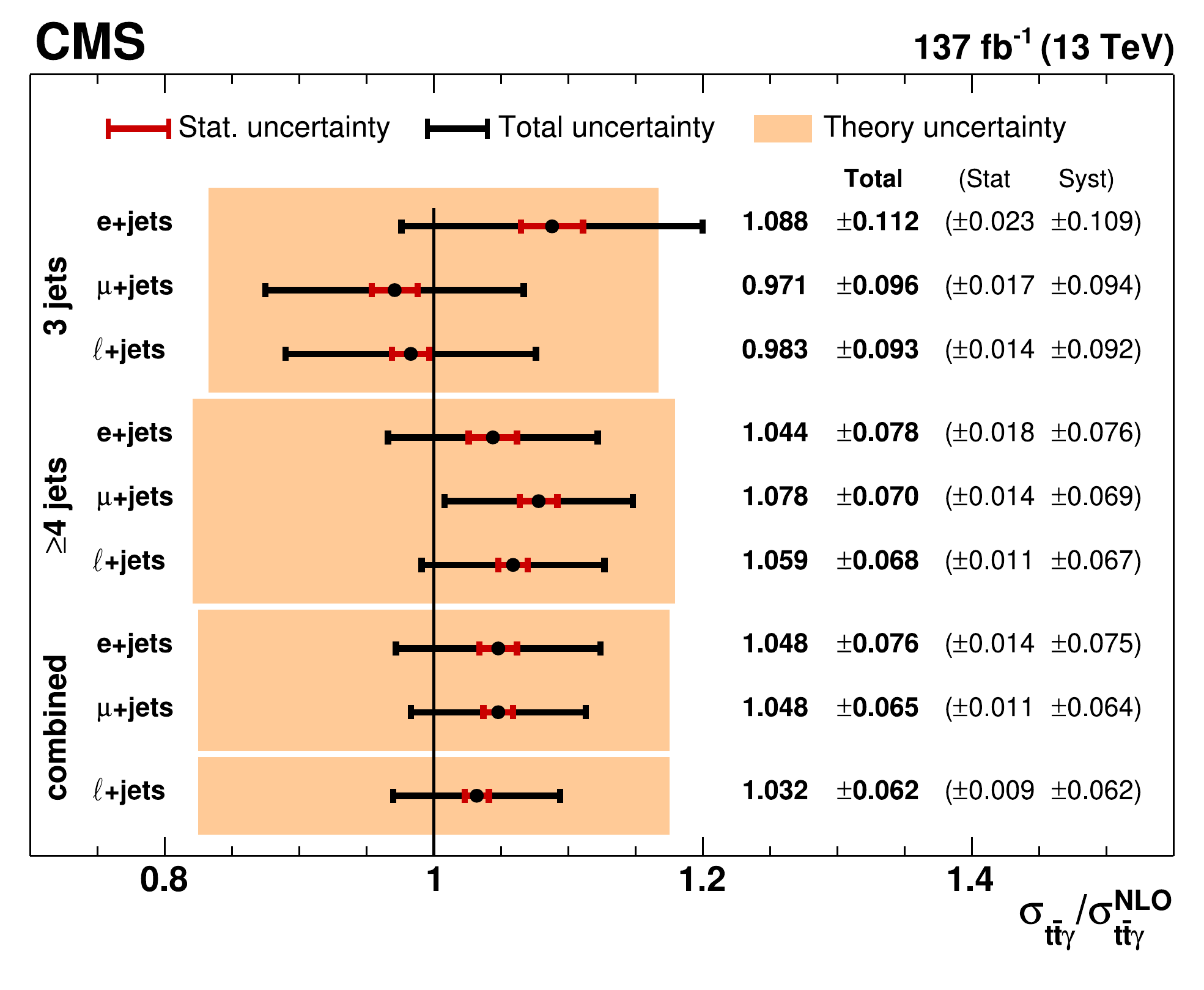}
  \caption{Summary of the measured cross section ratios with respect to the NLO cross section prediction for $\nJet=3$, $\geq$4, and combined signal regions in the electron channel, muon channel, and the combined single-lepton channel. The orange band indicates the theory uncertainty in the prediction. }
  \label{fig:summary_fit}
\end{figure}

The combined inclusive cross section of the $\nJet=3$ and $\geq$4 channels within the fiducial phase space is measured to be
\begin{equation}
\sigma(\ttg) = 798\pm 7\stat \pm 48 \syst\fb
\end{equation}
in good agreement with the SM expectation of $\sigma^\textrm{NLO}(\ttg)=773 \pm 135$\fb. 
The measured value of the signal strength modifier is 
\begin{equation}
r = 1.032\pm 0.009\stat \pm 0.062\syst. 
\end{equation}
A comparison of the measured cross sections and the SM prediction is shown in Fig.~\ref{fig:summary_fit}, providing also the measurements for different choices of \nJet and the lepton flavor.
For the latter results, the likelihood fit is performed separately in the corresponding set of signal regions and the full set of control regions. 

\subsection{Differential cross section measurement}\label{sec:results-diff}

The differential cross section is measured as a function of \ptG, \etaG, and \dRlg.
Results are obtained simultaneously for the electron and muon channels, the 3 jet and $\geq$4 jet bins, and for the three data-taking periods.
The binning in the SR3 and SR4p selections for the measurement of the differential distributions at the reconstruction level is shown in Table~\ref{tab:SR-binning}.

{\renewcommand{\arraystretch}{1.3}
\begin{table}[th!]
\topcaption{Binning choices in the differential measurements at the reconstruction level.}\label{tab:SR-binning}
\centering
\begin{tabular}{cl}
\ptG & 20, 35, 50, 65, 80, 100, 120, 140, 160, 200, 260, $\geq$320\GeV\\
\etaG & 0, 0.15, 0.30, 0.45, 0.60, 0.75, 0.90, 1.05, 1.20, 1.35, 1.4442\\
\dRlg & 0.4, 0.6, 0.8, 1.0, 1.2, 1.4, 1.6, 1.8, 2.0, 2.2, 2.4, 2.6, 2.8, $\geq$3.0 \\
\end{tabular}
\end{table}
}

As described in Section~\ref{sec:stat}, the same control regions are used for the inclusive and differential cross section measurements. 
The signal strength is left floating in the profile likelihood fit separately for each of the differential bins, the \nJet selection, the lepton flavor, and the data-taking period.
The procedure has been tested to reproduce ad-hoc modifications of the simulated signal prediction within the numerical accuracy. 
The fit is performed separately for each differential distribution. 

The distributions of the observables after background subtraction are further unfolded to the fiducial particle level phase space defined in Section~\ref{sec:sel_fiducial}.
The unfolded differential cross section is defined in the same phase space as the inclusive cross section reported above, \ie, in the phase space where the top quark pair is produced in association with a photon satisfying $\ptG>20\GeV$ and $\etaG<1.4442$. 
Signal events that are not generated within the fiducial region amount to 5--10\% and are subtracted based on simulation.
In the simulation, \ptG is taken as the transverse momentum after accounting for the effects of QCD and electroweak radiation. 

\begin{figure}[pt!]
    \centering
    \includegraphics[width=0.49\textwidth]{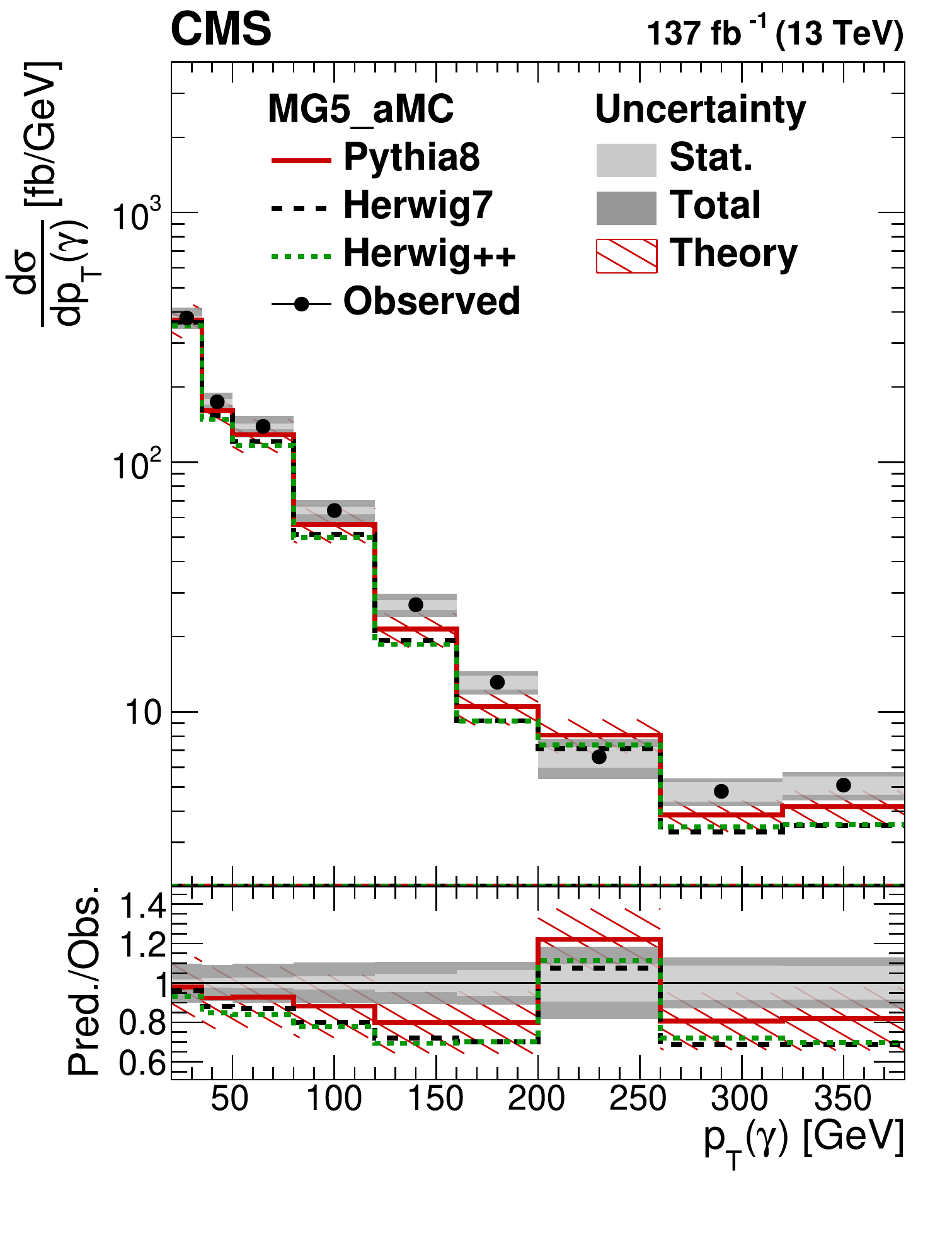}
    \includegraphics[width=0.49\textwidth]{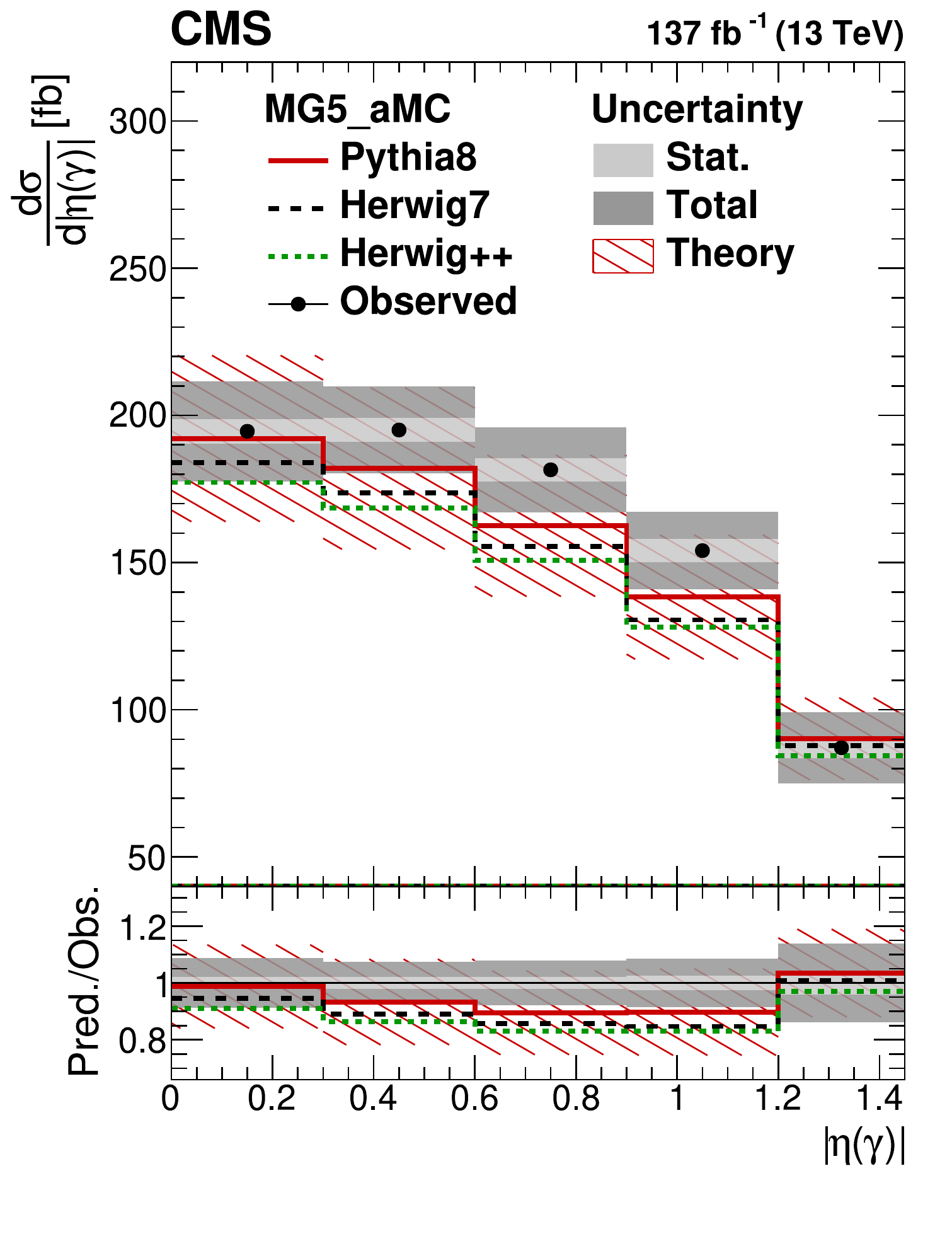}\\
    \includegraphics[width=0.49\textwidth]{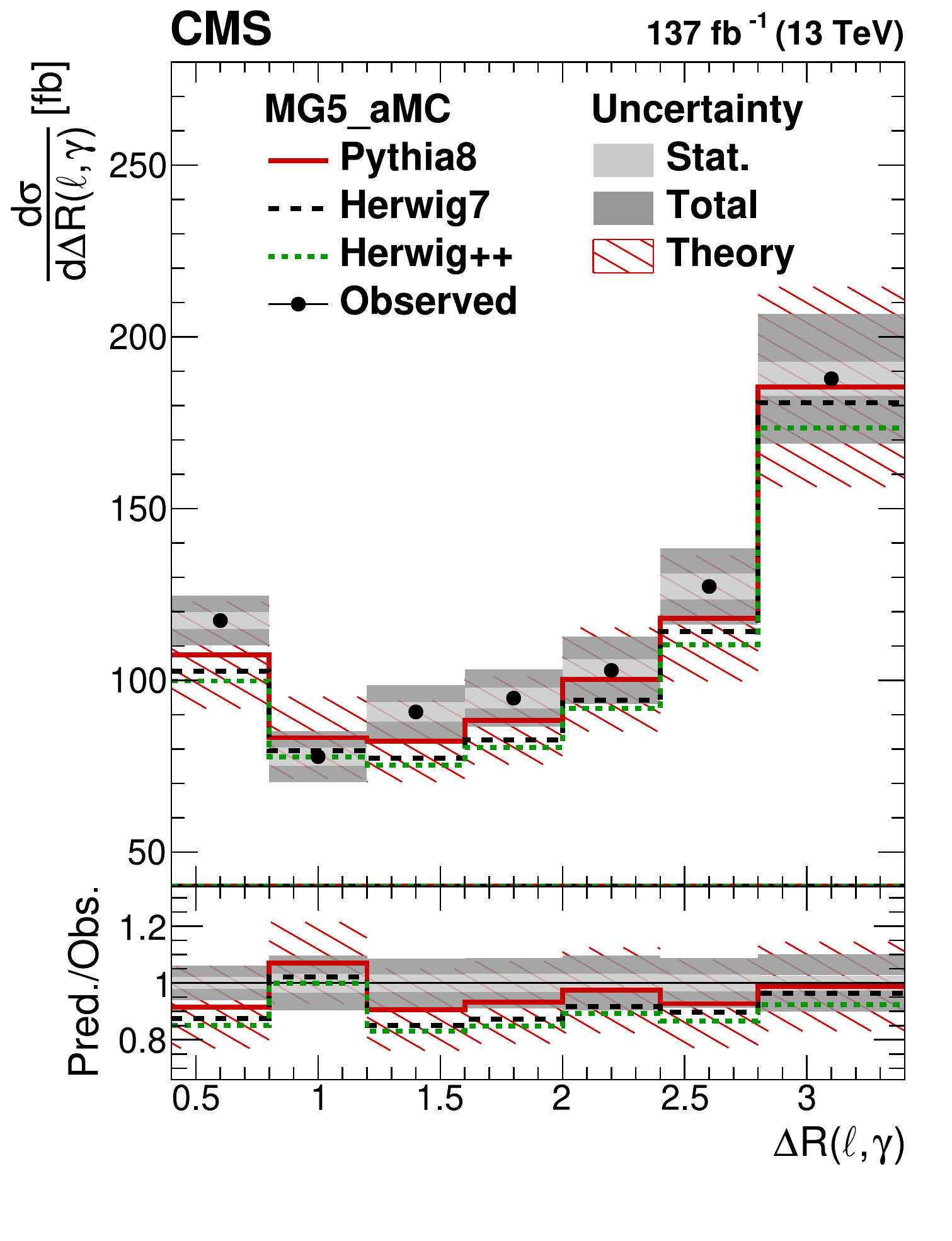}
  \caption{The unfolded differential cross sections for \ptG~(upper left), \etaG~(upper right), and \dRlg~(lower) compared with simulation obtained from the \MGvATNLO event generator interfaced to \PYTHIA (red, solid), \HERWIGSeven (black, dashed) and \HERWIGpp (green, dotted) for the parton shower and hadronization. 
    For \ptG and \dRlg, the last bin includes the overflow.
    The lower panel displays the ratio of simulation to the observation. The inner and outer bands show the statistical and total uncertainties, respectively.
   Photons radiated from leptons and satisfying $\dRlg>0.4$ are included in the signal and contribute significantly to the first bin of the differential  \dRlg cross section. 
}
  \label{fig:unfolding_output}
\end{figure}

\begin{figure}[pt!]
    \centering
    \includegraphics[width=0.49\textwidth]{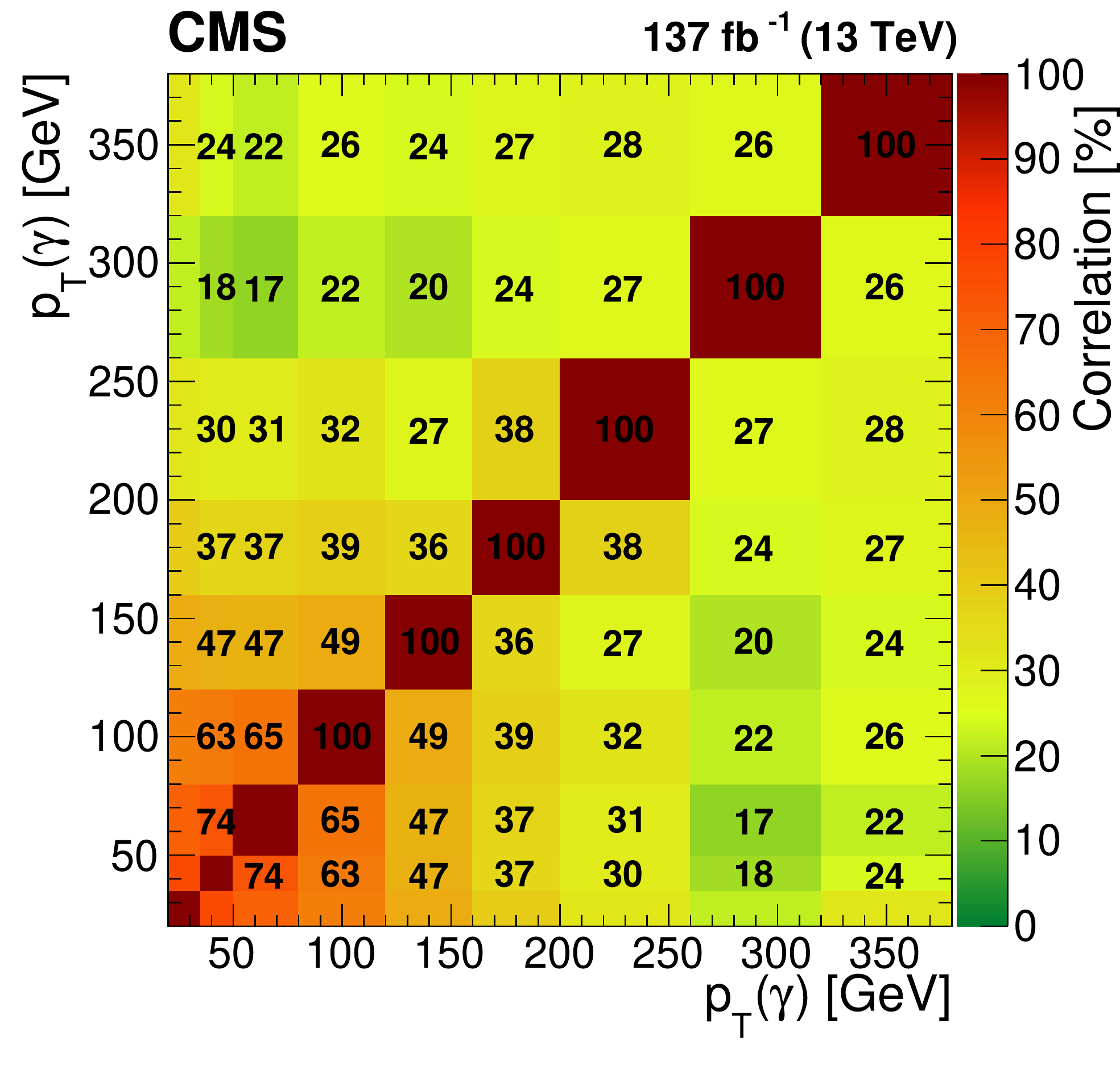}
    \includegraphics[width=0.49\textwidth]{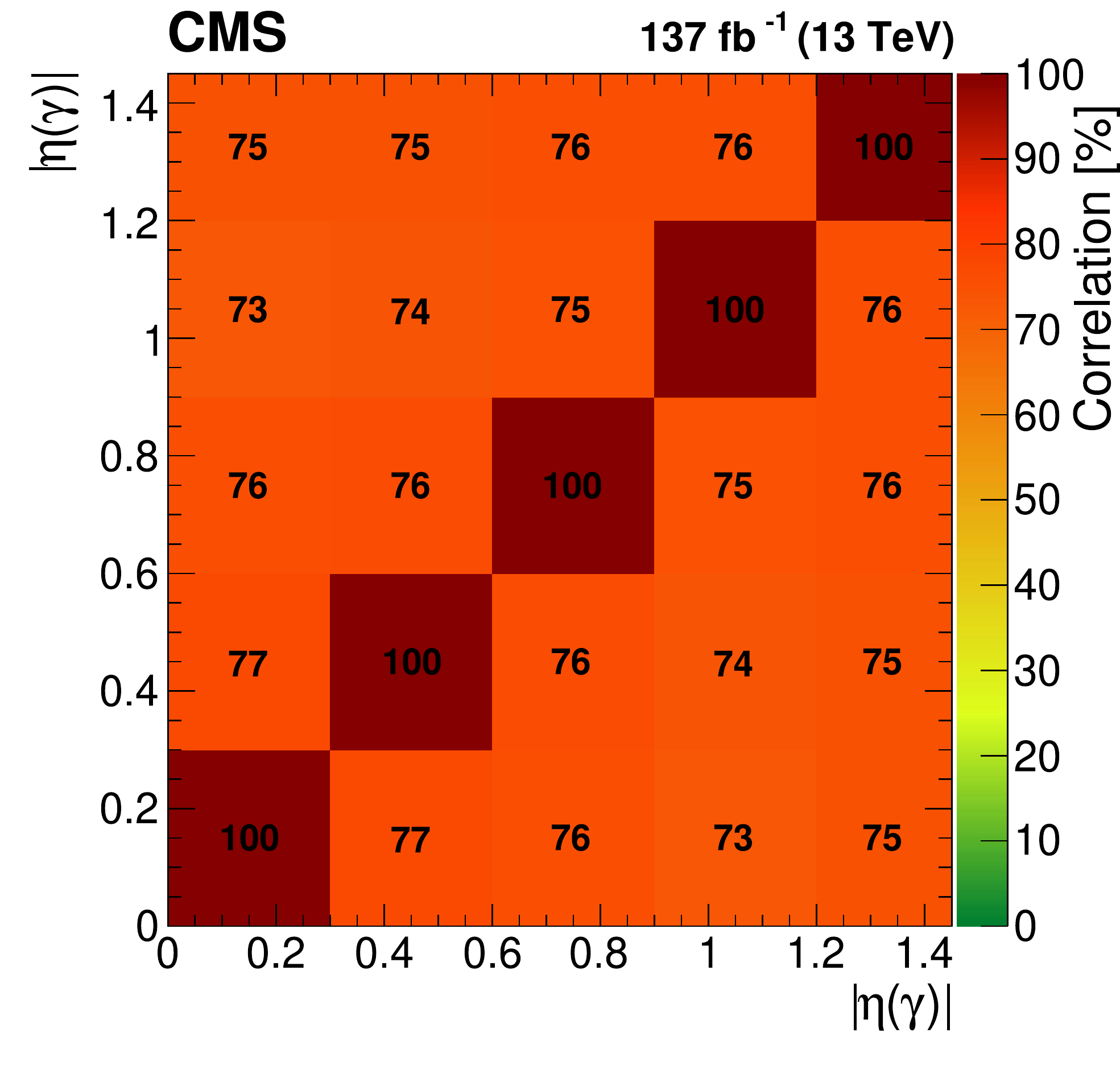}\\
    \includegraphics[width=0.49\textwidth]{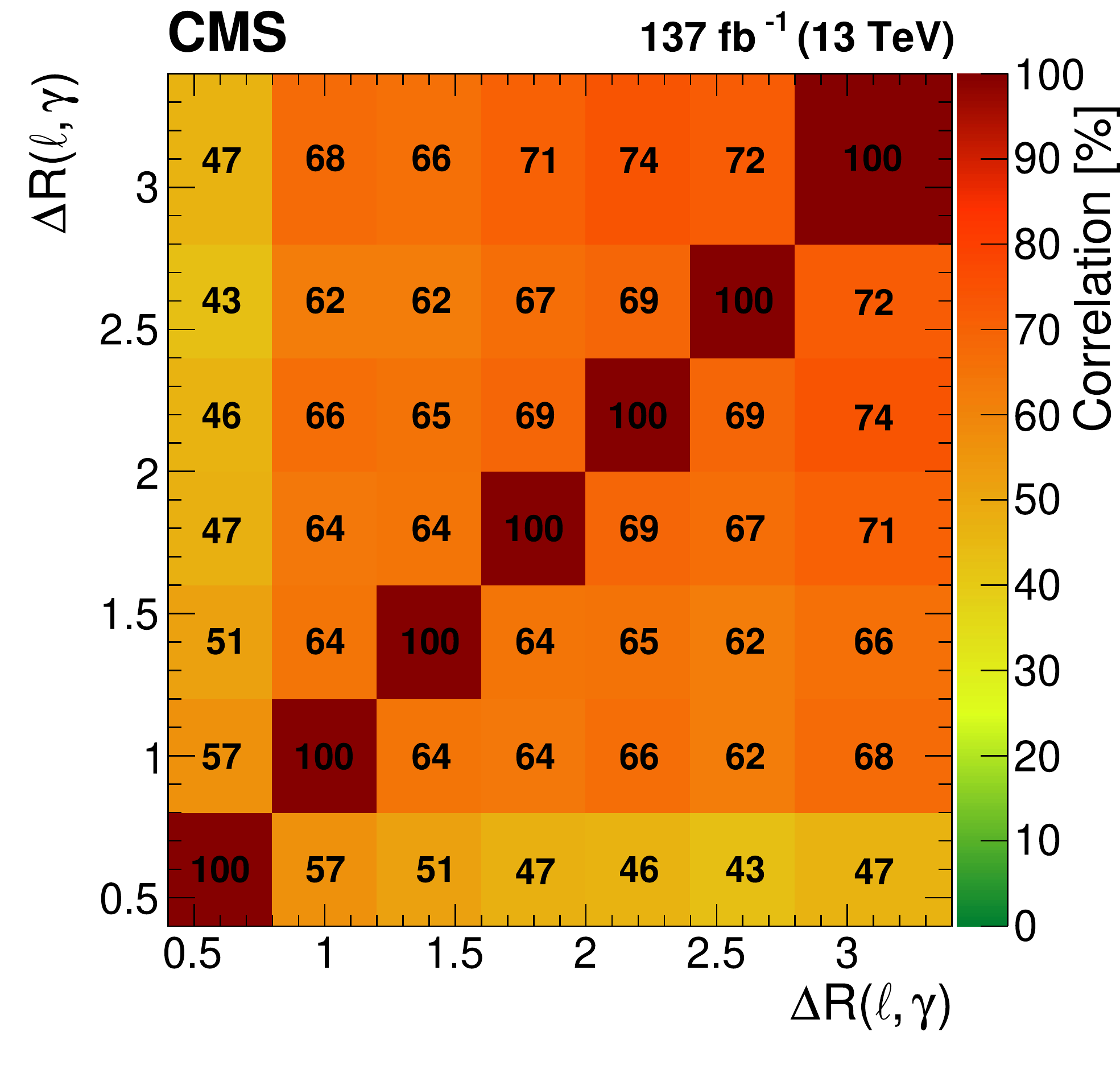}
  \caption{The correlation matrices of systematic uncertainties for the unfolded differential measurement for \ptG~(upper left), \etaG~(upper right), and \dRlg~(lower).}
  \label{fig:corrMatrices_diff}
\end{figure}

The \ttg \MGvATNLO MC sample is used to construct a response matrix that takes into account both detector response and acceptance corrections. 
The same corrections, SFs, and uncertainties as used in the inclusive cross section are applied. 
Because of the high momentum- and angular resolutions of photons and leptons, the fraction of events migrating from a specific momentum region at the particle level to another one at the reconstruction level is small for all unfolded distributions.
Under such conditions, and with the chosen bin size, no regularization term is required~\cite{Cowan:1998ji}.
The \textsc{TUnfold} package~\cite{Schmitt:2012kp} is used to obtain the results for the three measured observables using matrix inversion.
The binning in the fiducial region is chosen such that two bins at the reconstruction level correspond to one bin in the fiducial region for most cases.
This choice provides stability to the unfolding algorithm.
The linearity of the unfolding procedure is tested by unfolding suitably reweighted simulated reconstruction-level yields. 
Differences between the unfolded reweighted distributions and the distributions resulting from the reweighting applied at the fiducial level are found to be negligible.

Uncertainties in the estimated signal yields are propagated through the unfolding procedure, including the effects on the response matrix.
Experimental uncertainties from the detector response and efficiency, such as the photon identification, JES, and \PQb~tagging uncertainties, are applied as a function of the reconstructed observable. 
The differential cross sections, obtained by this procedure, are shown in Fig.~\ref{fig:unfolding_output}.
It includes a comparison with simulation obtained from \MGvATNLO interfaced to \HERWIGpp~\cite{Bellm:2015jjp}~v2.7.1 with the EE5C tune~\cite{Khachatryan:2015pea} and to \HERWIGSeven~v7.1.4 with the CH3 tune~\cite{Sirunyan:2020pqv} for the parton shower and hadronization. 
The inclusive fiducial cross section predicted by \HERWIGpp~(\HERWIGSeven) is 8.3\%~(5.4\%) lower than for the nominal simulation.

The bin efficiency, defined as the fraction of generated events that are reconstructed in the corresponding bins at reconstruction level, is in the range of 20--30\%.
The bin purity, defined as the fraction of reconstructed events that originate from the corresponding bin at the particle level, is in the range of 85--90\%.
For $\ptG>120\GeV$, the uncertainties in the JES, the photon identification efficiency, and the color reconnection modeling are the largest sources of systematic uncertainty. 
The correlation matrices of the systematic uncertainties for the unfolded differential measurements are shown in Fig.~\ref{fig:corrMatrices_diff}.
The correlations are lower in the tail of \ptG due to larger statistical uncertainties in the simulation.
The first bin of the \dRlg measurement is less affected by uncertainties in the normalization of backgrounds, resulting in slightly lower correlations in this case.
All correlations from statistical uncertainties originating from the data are below 7\%.
Including the uncertainty in the fiducial signal cross section, we perform a compatibility test of the unfolded distribution and the nominal prediction. 
The corresponding $\chi^2$ test statistic evaluates to 12.0 with 9 degrees of freedom~(dof) for the \ptG distribution, 5.2 with 5 dof for \etaG, and 6.3 with 7 dof for \dRlg.

\subsection{Effective field theory interpretation}\label{sec:eft}

Many BSM models predict anomalous couplings of the top quark to the electroweak gauge bosons~\cite{Hollik:1998vz,Agashe:2006wa,Kagan:2009bn,Ibrahim:2010hv,Ibrahim:2011im,Grojean:2013qca,Richard:2013pwa}.
The differential cross section measurement is interpreted at the reconstruction level in SM-EFT in the Warsaw basis~\cite{Grzadkowski:2010es}, formed by 59 baryon number conserving dimension-six Wilson coefficients.
Among them, 15 are relevant for top quark interactions~\cite{Zhang:2010dr}. 
Anomalous interactions between the top quark and the gluon (chromomagnetic and chromoelectric dipole moment interactions) are tightly constrained by the {\ttbar}+jets measurements~\cite{Sirunyan:2018ucr,CMS:2018jcg}.
Similarly, the modification of the $\PW\PQt\Pb$ vertex is best constrained by measurements of the \PW~helicity fractions in top quark pair production~\cite{Khachatryan:2016fky} and in $t$-channel single top quark production~\cite{AguilarSaavedra:2010nx}.

The Wilson coefficients in the Warsaw basis inducing electroweak dipole moments are denoted by $C_{\PQu\cmsSymbolFace{B}}^{(33)}$ and $C_{\PQu\PW}^{(33)}$~\cite{AguilarSaavedra:2018nen}.
The SM gauge symmetry provides the \ttZ and the \ttg final states with complementary constraining power~\cite{Baur:2004uw,Bouzas:2012av,Schulze:2016qas,Rontsch:2015una}. 
The linear relations
\begin{equation*}
  \begin{aligned}
\ctZ  &= \mathrm{Re}\left( -\sinw C_{\PQu\cmsSymbolFace{B}}^{(33)} + \cosw C_{\PQu\PW}^{(33)}\right), \\
\ctZI &= \mathrm{Im}\left( -\sinw C_{\PQu\cmsSymbolFace{B}}^{(33)} + \cosw C_{\PQu\PW}^{(33)}\right), \\
\ctA  &= \mathrm{Re}\left(\cosw C_{\PQu\cmsSymbolFace{B}}^{(33)} + \sinw C_{\PQu\PW}^{(33)}\right), \\
\ctAI &= \mathrm{Im}\left(\cosw C_{\PQu\cmsSymbolFace{B}}^{(33)} + \sinw C_{\PQu\PW}^{(33)}\right),
  \end{aligned}
\end{equation*}
express the modifications of the \ttZ interaction vertex, \ctZ and \ctZI,  and of the \ttg interaction vertex, \ctA and \ctAI, in the Warsaw basis. 
The constraint $C_{\PQu\PW}^{(33)}=0$ ensures a SM $\PW\PQt\PQb$ vertex.
Under this assumption, \ctZ(\ctZI) and \ctA(\ctAI) are dependent and we choose the former to parametrize the BSM hypothesis.

The spectrum of \ptG is a sensitive probe to such modifications.
Other observables, \eg, \etaG or \dRlg, are found to be largely insensitive. 
Wilson coefficients that are not considered in this work are kept at their SM values and the SM-EFT expansion parameter is set to a mass scale $\Lambda=1\TeV$.
Using the SM-EFT parametrization from Ref.~\cite{AguilarSaavedra:2018nen}, simulated samples at the particle level are produced with nonzero values of the Wilson coefficients \ctZ and \ctZI.
The \ttg signal process and all background processes affected by \ctZ or \ctZI at the ME level are included in the simulation. 
These samples are used to reweight the nominal simulation in the fiducial phase space using the quadratic parametrization detailed in Ref.~\cite{Sirunyan:2020tqm}.
The reweighting procedure is validated at the reconstruction level with a reduced set of statistically independent samples for nonzero values of \ctZ and \ctZI and excellent agreement is found. 

The SR3 and SR4p signal regions and the \ptG boundaries defining the bins in Table~\ref{tab:SR-binning} are used to construct a binned likelihood function $L(\theta)$ as a product of Poisson probabilities from the yields in the signal and control regions.
The nuisance parameters are labeled by $\theta$ and the profile likelihood ratio $q=-2\ln(L(\hat{\theta}, \vec C)/L(\hat{\theta}_{\text{max}}))$ is the test statistic.
Here, $\hat{\theta}$ is the set of nuisance parameters maximizing the likelihood function at a BSM point defined by the Wilson coefficients collectively denoted by $\vec C$.
In the denominator, $\hat{\theta}_\text{max}$ maximizes the likelihood function in the BSM parameter space.
The \ttg signal is normalized according to the SM expectation at NLO in QCD and its uncertainty is included as a nuisance.

\begin{figure}[tph]
  \centering
    \includegraphics[width=0.49\textwidth]{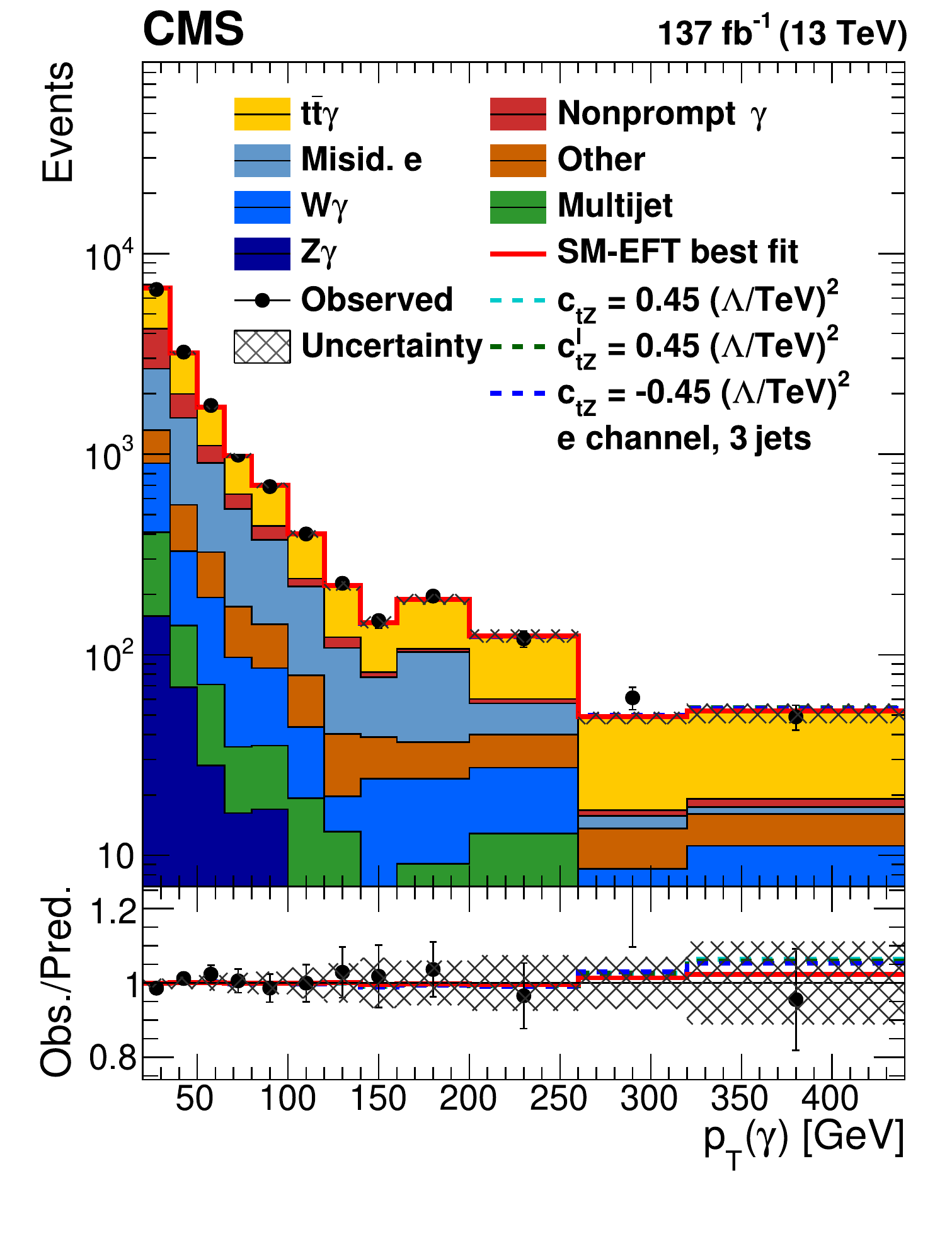}
    \includegraphics[width=0.49\textwidth]{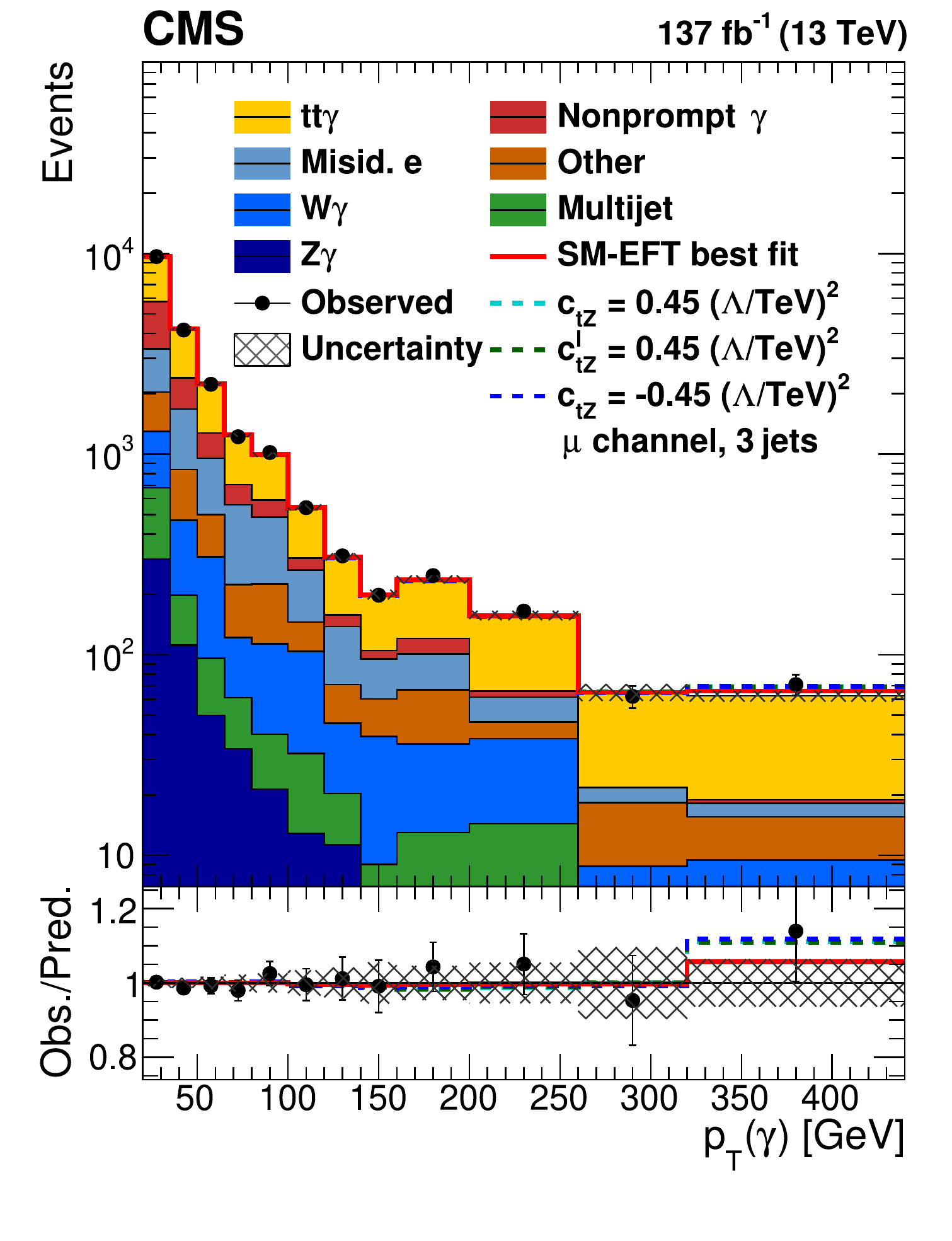}\\
    \includegraphics[width=0.49\textwidth]{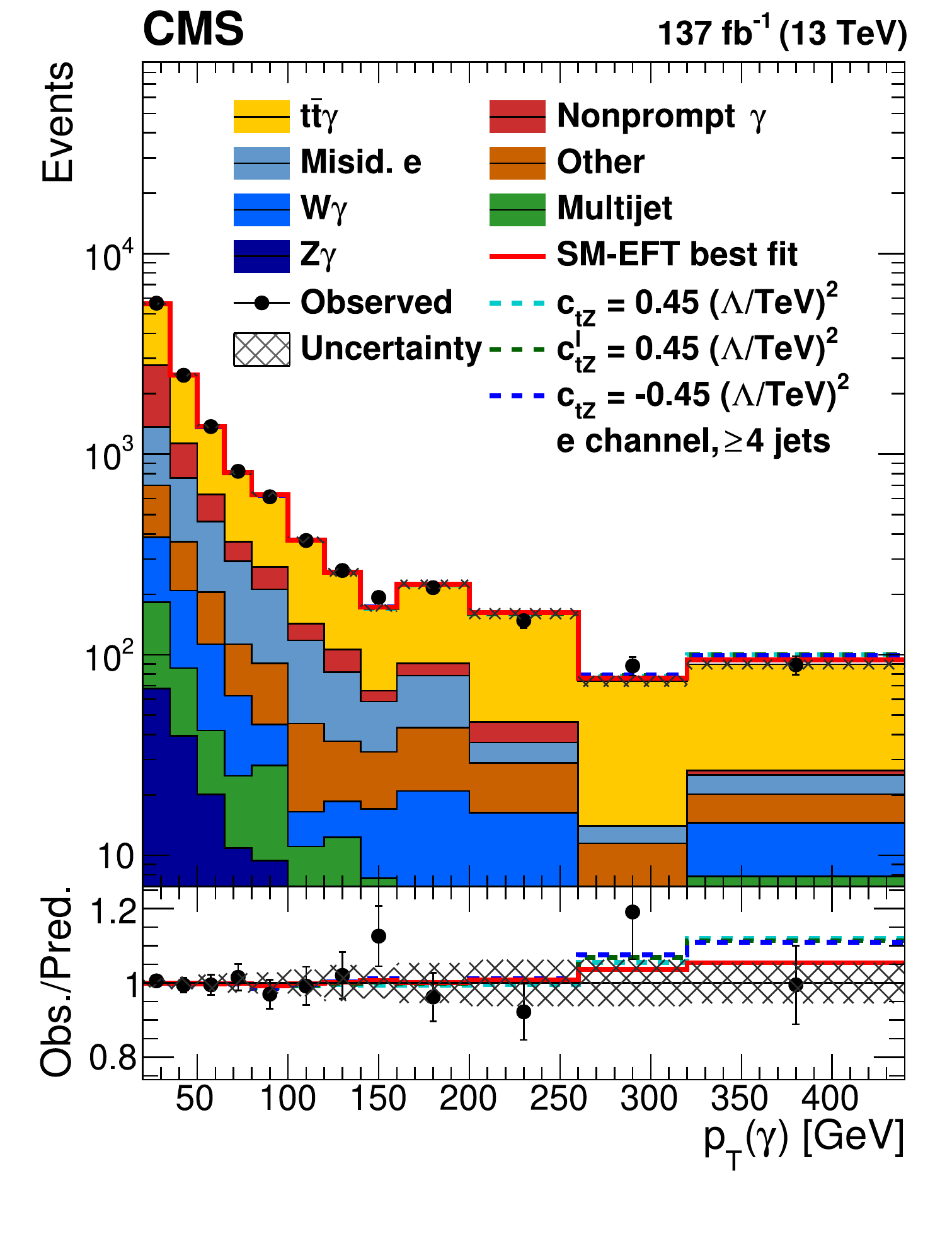}
    \includegraphics[width=0.49\textwidth]{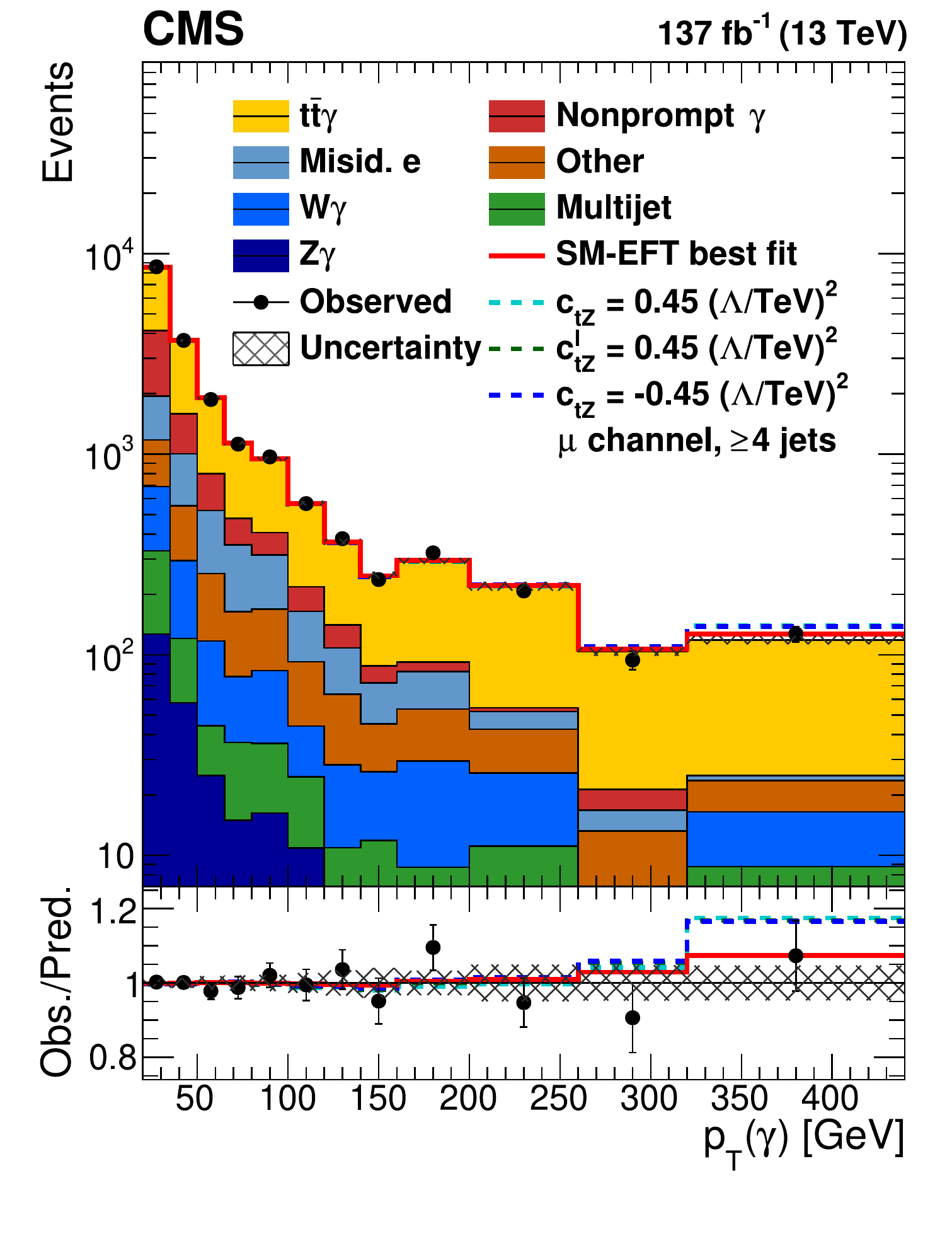}
  \caption{ The observed (points) and predicted (shaded histograms) post-fit yields for the combined Run~2 data set in the SR3~(upper) and SR4p~(lower) signal regions for the electron~(left) and muon channel~(right). 
The vertical bars on the points give the statistical uncertainties in the data. 
The lower panel displays the ratio of the data to the predictions and the hatched regions show the total uncertainty. 
The solid line shows the SM-EFT best fit prediction and the dashed lines show different predictions for nonzero Wilson coefficients, $\ctZ=0.45$~(light blue), $\ctZI=0.45$~(green), and $\ctZ=-0.45$~(dark blue), where $\Lambda$ is set to 1\TeV.}
  \label{fig:EFT_pt}
\end{figure}

\begin{figure}[tph]
  \centering
    \includegraphics[width=0.49\textwidth]{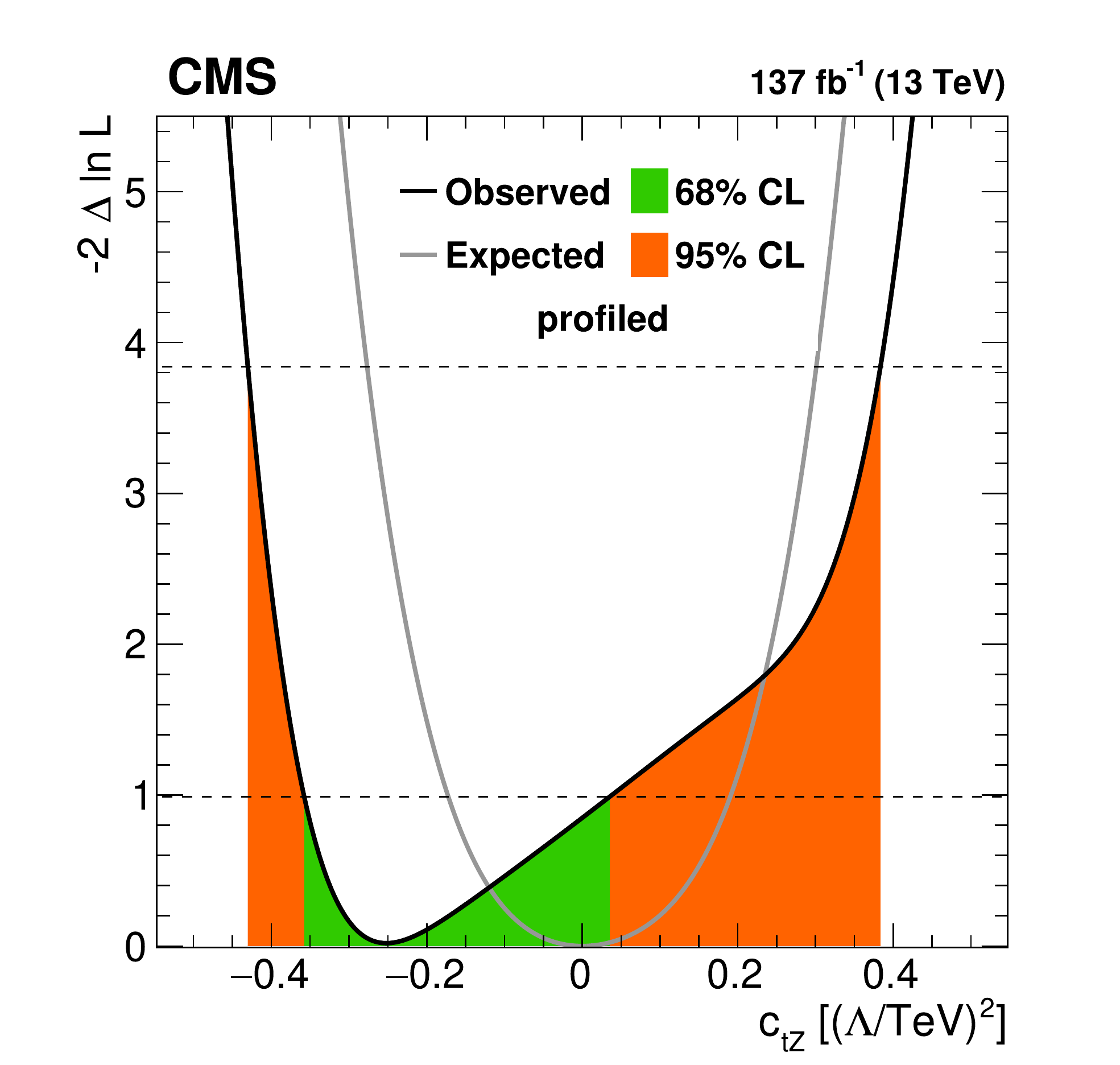}
    \includegraphics[width=0.49\textwidth]{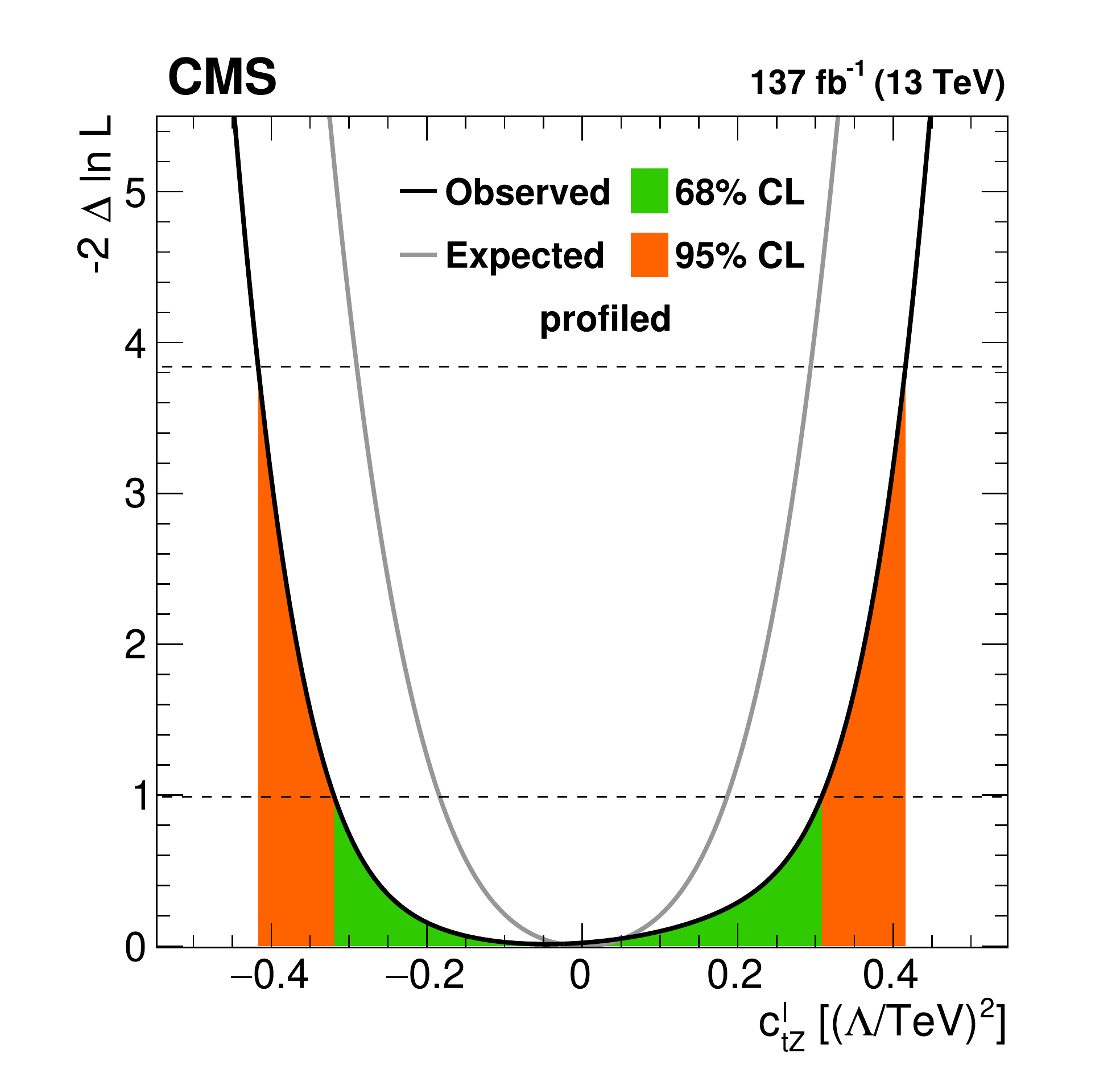}\\
    \includegraphics[width=0.49\textwidth]{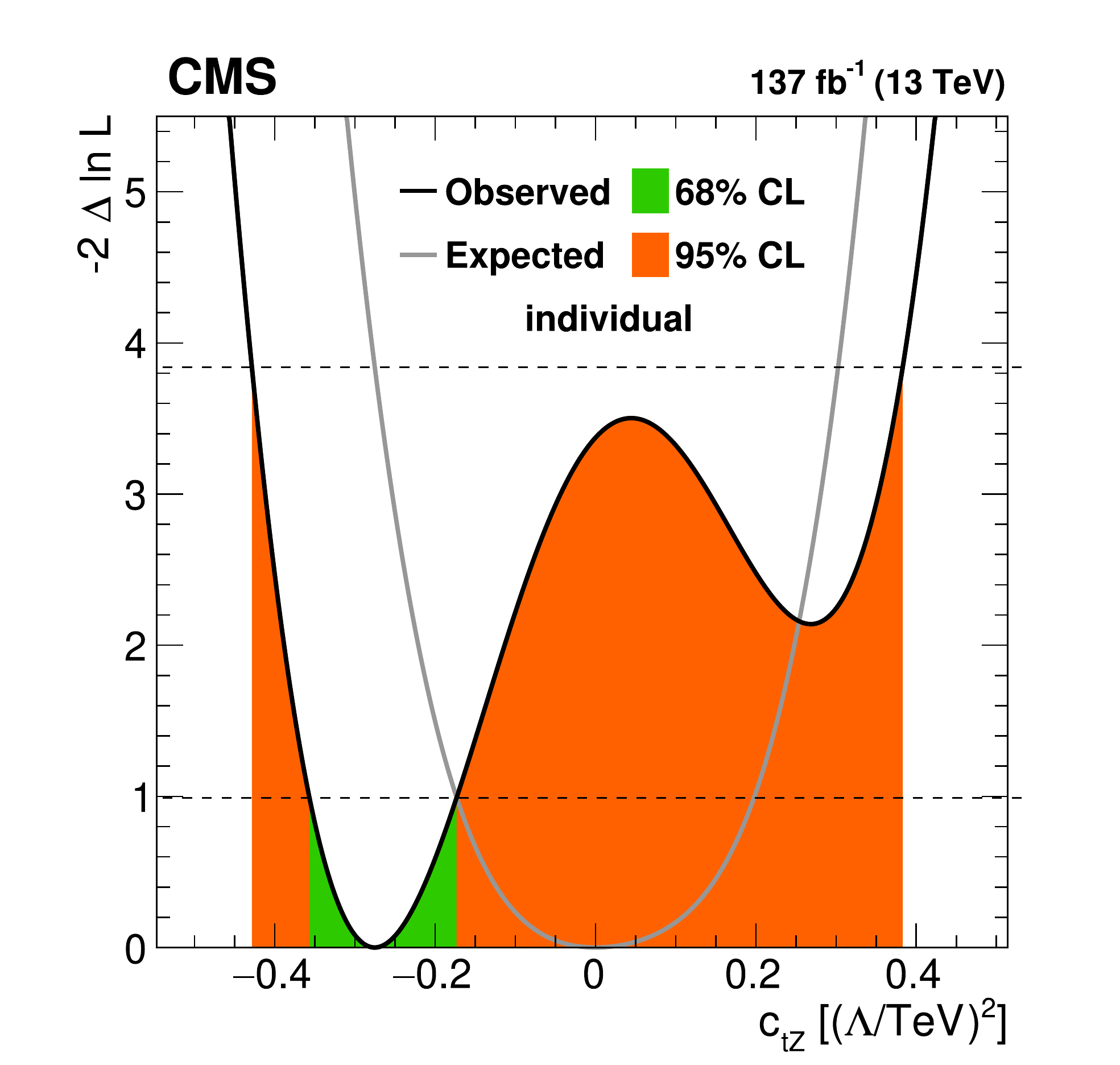}
    \includegraphics[width=0.49\textwidth]{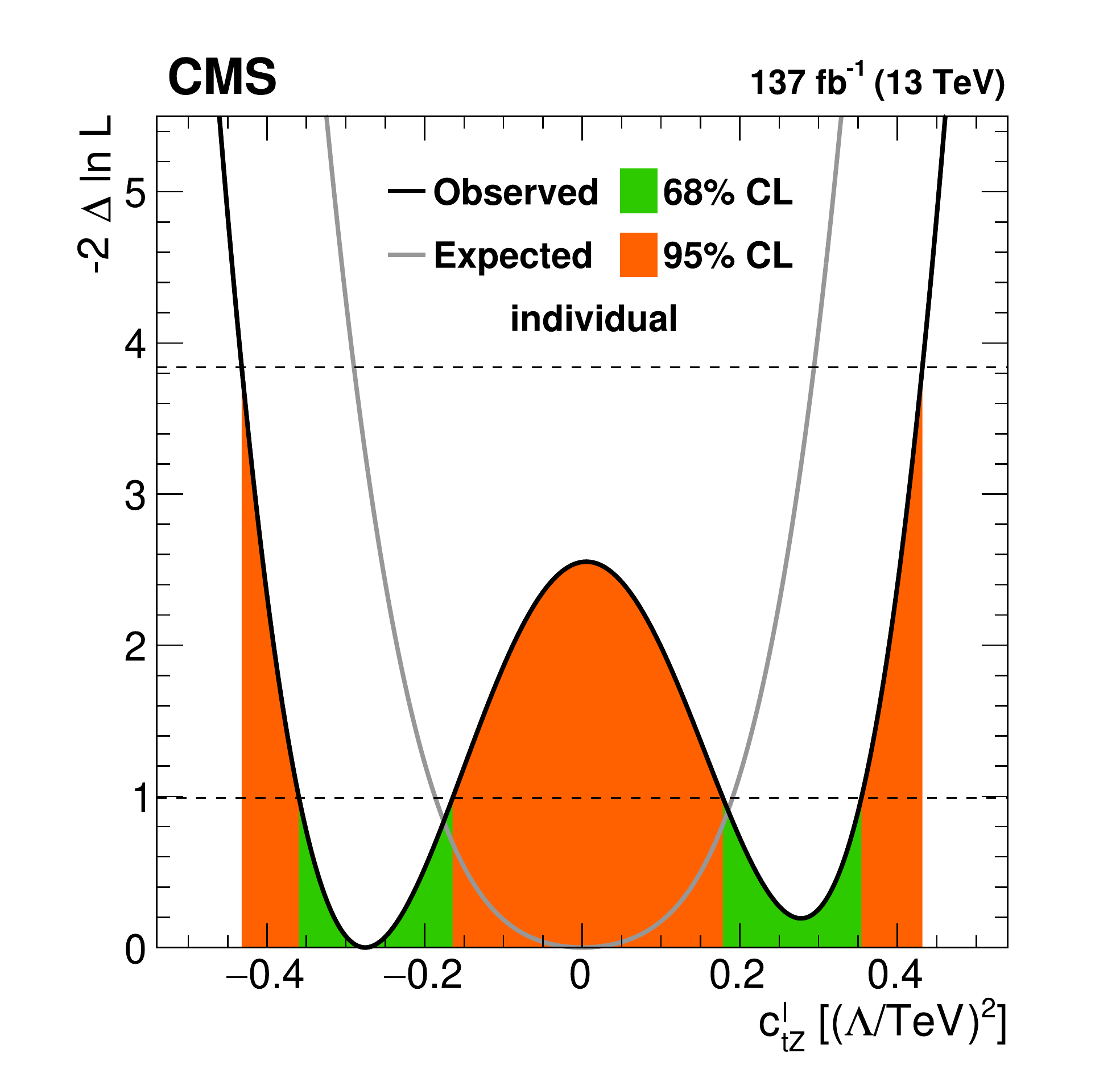}
  \caption{Results of the one-dimensional scans of the Wilson coefficients \ctZ~(left) and \ctZI~(right). 
In the upper row, the other Wilson coefficient is profiled, while in the lower row it is set to zero.
The green and orange bands indicate the 68 and 95\% \CL contours on the Wilson coefficients, respectively. }
  \label{fig:EFT_results}
\end{figure}

\begin{figure}[tph]
  \centering
    \includegraphics[width=0.65\textwidth]{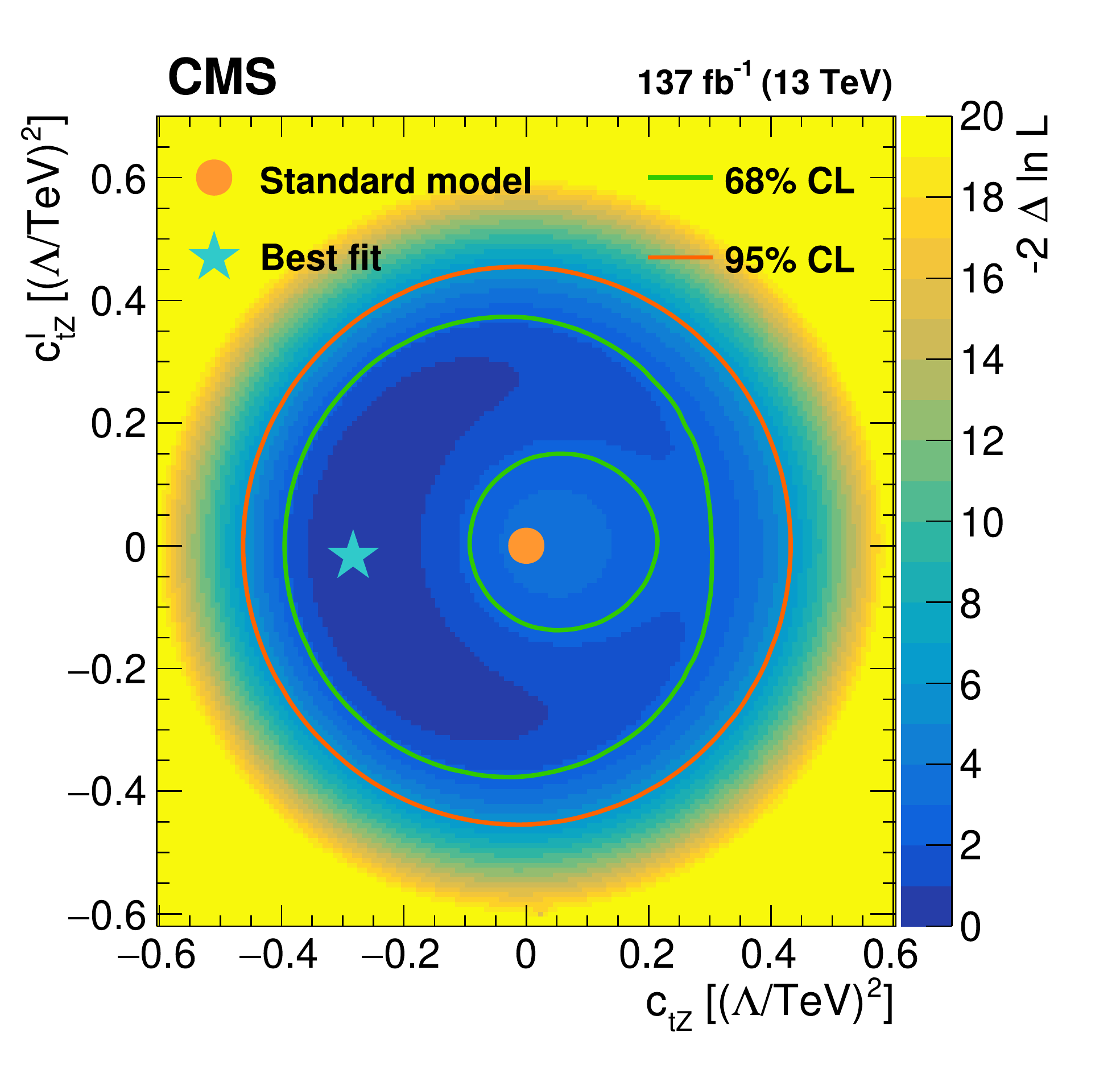}
  \caption{Result of the two-dimensional scan of the Wilson coefficients \ctZ and \ctZI.
The shading quantified by the color scale on the right reflects the negative log-likelihood ratio with respect to the best fit value that is designated by the star.
The green and orange lines indicate the 68 and 95\% \CL contours from the fit, respectively. 
The allowed areas are those between the two green contours and that inside the orange contour. 
The dot shows the SM prediction. 
}
  \label{fig:EFT_results_2D}
\end{figure}

Figure~\ref{fig:EFT_pt} shows the result of the fit for the SR3 and SR4p signal regions and separately for each lepton flavor. 
No deviations from the SM expectations are observed.
The best fit point is found at ($\ctZ$, $\ctZI$) = ($-0.28$, $-0.02$) and the corresponding spectrum is overlaid together with the ones from several other choices for nonzero values of the Wilson coefficients. 
Figure~\ref{fig:EFT_results} displays the one-dimensional scans of the coefficients.
In the upper row, one Wilson coefficient is scanned, while the other is profiled.
The lower row shows the scans, where the second Wilson coefficient is set to zero.
The second local minima in the scans of the log-likelihood as a function of \ctZ and \ctZI, visible in Fig.~\ref{fig:EFT_results}~(lower row), is the result of a mild tension with the SM hypothesis in conjunction with the similarity of the predictions for Wilson coefficients with opposite sign. 
The corresponding one-dimensional intervals at 68 and 95\% confidence interval (\CL) are listed in Table~\ref{tab:eftresults} and are more stringent than previous limits obtained from \ttZ final states~\cite{CMS:2019too,Aaboud:2019njj}.
Models with nonzero electroweak dipole moments predict a harder \ptG spectrum that is not observed in data.
Figure~\ref{fig:EFT_results_2D} shows the best fit result in the two-dimensional plane spanned by \ctZ and \ctZI and the log-likelihood scan.
The SM prediction is within the 95\% \CL of the best fit value of the \ctZ and \ctZI coefficients.

\begin{figure}[tph]
  \centering
    \includegraphics[width=0.55\textwidth]{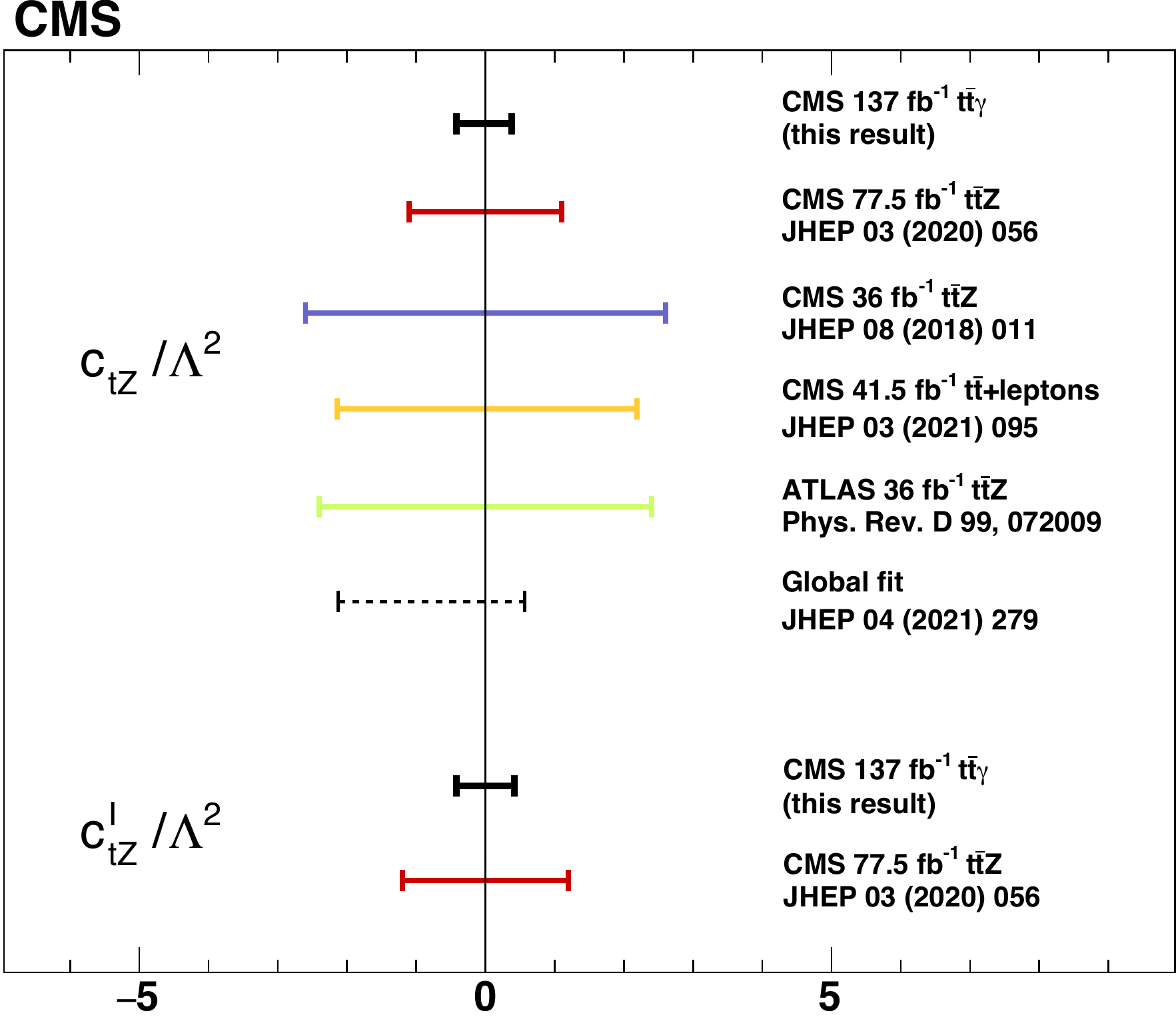}
  \caption{
The observed 95\% \CL intervals for the Wilson coefficients from this measurement with the other Wilson coefficient set to zero, the previous CMS results based on the inclusive~\cite{Sirunyan:2017uzs} and differential~\cite{CMS:2019too} \ttZ cross section measurement, a CMS result based on \ttbar in final states with additional leptons~\cite{Sirunyan:2020tqm}, and the most recent ATLAS result~\cite{Aaboud:2019njj}.
The result of a global SM-EFT analysis, including results from Ref.~\cite{CMS:2019too}, is also shown~\cite{Ellis:2020unq}.
The vertical line displays the SM prediction.
}
  \label{fig:limits}
\end{figure}

In Fig.\ref{fig:limits}, the 95\% \CL intervals are compared with the previous CMS results based on the inclusive~\cite{Sirunyan:2017uzs} and differential~\cite{CMS:2019too} \ttZ cross section measurement, a CMS result based on \ttbar in final states with additional leptons~\cite{Sirunyan:2020tqm}, and the most recent ATLAS result~\cite{Aaboud:2019njj}~(lower).
The result of a global SM-EFT analysis, including results from Ref.~\cite{CMS:2019too}, is also shown~\cite{Ellis:2020unq}.
The present result improves upon the previous constraints by about a factor~of~2.5.

{\renewcommand{\arraystretch}{1.3}
\begin{table}[tph]
  \topcaption{Summary of the one-dimensional intervals at 68 and 95\% \CL. }
\centering
  \begin{tabular}{lllcc }
                            & \multicolumn{2}{c}{\multirow{2}{*}{Wilson coefficient}} & 68\% \CL interval & 95\% \CL interval \\
                            &       & & $(\Lambda/\TeV)^{2}$& $(\Lambda/\TeV)^{2}$ \\
\hline
  \multirow{4}{*}{\rotatebox{90}{Expected}} & \multirow{2}{*}{\ctZ}  & $\ctZI=0$    & [$-0.19$, $0.20$]                & [$-0.29$, $0.31$] \\
                            &                        & profiled      & [$-0.19$, $0.20$]                & [$-0.29$, $0.31$] \\[5pt]
                            & \multirow{2}{*}{\ctZI} & $\ctZ=0$     & [$-0.20$, $0.20$]                & [$-0.30$, $0.30$] \\
                            &                        & profiled      & [$-0.20$, $0.20$]                & [$-0.30$, $0.30$] \\[15pt]
  \multirow{4}{*}{\rotatebox{90}{Observed}} & \multirow{2}{*}{\ctZ }  & $\ctZI=0$    & [$-0.36$, $-0.17$]               & [$-0.43$, $0.38$] \\
                            &                        &  profiled     & [$-0.36$, $0.04$]                & [$-0.43$, $0.38$] \\[5pt]
                            & \multirow{2}{*}{\ctZI} & $\ctZ=0$     & [$-0.36$, $-0.16$], [$0.18$, $0.35$] & [$-0.43$, $0.43$] \\
                            &                        & profiled      & [$-0.32$, $0.31$]                & [$-0.42$, $0.42$] 
  \end{tabular}
  \label{tab:eftresults}
\end{table}
}

\section{Summary}\label{sec:summary}

A measurement of the cross section for the top quark pair production in association with a photon using a data sample of proton-proton collisions at $\sqrt{s}=13\TeV$, corresponding to an integrated luminosity of 137\fbinv, collected with the CMS detector at the LHC has been presented. 
It is the first result of the CMS Collaboration on measurements in the \ttg final state using 13\TeV data.
The analysis has been performed in the single-lepton channel with events with exactly three and four or more jets among which at least one is \PQb~tagged.
Background components with misidentified electrons, photons originating in the hadronization of jets, the multijet component, and prompt photons from the \WGamma and \ZGamma processes are estimated from data.
The measured inclusive cross section in a fiducial region with photon transverse momentum $\ptG>20\GeV$ and jet multiplicity greater than three is measured to be $798\pm 7\stat \pm 48 \syst\fb$, in good agreement with the standard model prediction at next-to-leading order in quantum chromodynamics.

Differential cross sections for \ptG and absolute value of the photon pseudorapidity, as well as for the angular separation of the lepton and the photon, have been measured and unfolded to particle level in the same fiducial volume.
The comparison to simulation was performed using different showering algorithms.
The measurements are also interpreted in terms of limits on the Wilson coefficients in the context of the standard model effective field theory.
The confidence intervals for the Wilson coefficients \ctZ and \ctZI are the most stringent to date.

\clearpage

\begin{acknowledgments}
  We congratulate our colleagues in the CERN accelerator departments for the excellent performance of the LHC and thank the technical and administrative staffs at CERN and at other CMS institutes for their contributions to the success of the CMS effort. In addition, we gratefully acknowledge the computing centers and personnel of the Worldwide LHC Computing Grid and other centers for delivering so effectively the computing infrastructure essential to our analyses. Finally, we acknowledge the enduring support for the construction and operation of the LHC, the CMS detector, and the supporting computing infrastructure provided by the following funding agencies: BMBWF and FWF (Austria); FNRS and FWO (Belgium); CNPq, CAPES, FAPERJ, FAPERGS, and FAPESP (Brazil); MES (Bulgaria); CERN; CAS, MoST, and NSFC (China); MINCIENCIAS (Colombia); MSES and CSF (Croatia); RIF (Cyprus); SENESCYT (Ecuador); MoER, ERC PUT and ERDF (Estonia); Academy of Finland, MEC, and HIP (Finland); CEA and CNRS/IN2P3 (France); BMBF, DFG, and HGF (Germany); GSRT (Greece); NKFIA (Hungary); DAE and DST (India); IPM (Iran); SFI (Ireland); INFN (Italy); MSIP and NRF (Republic of Korea); MES (Latvia); LAS (Lithuania); MOE and UM (Malaysia); BUAP, CINVESTAV, CONACYT, LNS, SEP, and UASLP-FAI (Mexico); MOS (Montenegro); MBIE (New Zealand); PAEC (Pakistan); MSHE and NSC (Poland); FCT (Portugal); JINR (Dubna); MON, RosAtom, RAS, RFBR, and NRC KI (Russia); MESTD (Serbia); SEIDI, CPAN, PCTI, and FEDER (Spain); MOSTR (Sri Lanka); Swiss Funding Agencies (Switzerland); MST (Taipei); ThEPCenter, IPST, STAR, and NSTDA (Thailand); TUBITAK and TAEK (Turkey); NASU (Ukraine); STFC (United Kingdom); DOE and NSF (USA).
  
  \hyphenation{Rachada-pisek} Individuals have received support from the Marie-Curie program and the European Research Council and Horizon 2020 Grant, contract Nos.\ 675440, 724704, 752730, 765710 and 824093 (European Union); the Leventis Foundation; the Alfred P.\ Sloan Foundation; the Alexander von Humboldt Foundation; the Belgian Federal Science Policy Office; the Fonds pour la Formation \`a la Recherche dans l'Industrie et dans l'Agriculture (FRIA-Belgium); the Agentschap voor Innovatie door Wetenschap en Technologie (IWT-Belgium); the F.R.S.-FNRS and FWO (Belgium) under the ``Excellence of Science -- EOS" -- be.h project n.\ 30820817; the Beijing Municipal Science \& Technology Commission, No. Z191100007219010; the Ministry of Education, Youth and Sports (MEYS) of the Czech Republic; the Deutsche Forschungsgemeinschaft (DFG), under Germany's Excellence Strategy -- EXC 2121 ``Quantum Universe" -- 390833306, and under project number 400140256 - GRK2497; the Lend\"ulet (``Momentum") Program and the J\'anos Bolyai Research Scholarship of the Hungarian Academy of Sciences, the New National Excellence Program \'UNKP, the NKFIA research grants 123842, 123959, 124845, 124850, 125105, 128713, 128786, and 129058 (Hungary); the Council of Science and Industrial Research, India; the Latvian Council of Science; the Ministry of Science and Higher Education and the National Science Center, contracts Opus 2014/15/B/ST2/03998 and 2015/19/B/ST2/02861 (Poland); the National Priorities Research Program by Qatar National Research Fund; the Ministry of Science and Higher Education, project no. 0723-2020-0041 (Russia); the Programa Estatal de Fomento de la Investigaci{\'o}n Cient{\'i}fica y T{\'e}cnica de Excelencia Mar\'{\i}a de Maeztu, grant MDM-2015-0509 and the Programa Severo Ochoa del Principado de Asturias; the Thalis and Aristeia programs cofinanced by EU-ESF and the Greek NSRF; the Rachadapisek Sompot Fund for Postdoctoral Fellowship, Chulalongkorn University and the Chulalongkorn Academic into Its 2nd Century Project Advancement Project (Thailand); the Kavli Foundation; the Nvidia Corporation; the SuperMicro Corporation; the Welch Foundation, contract C-1845; and the Weston Havens Foundation (USA).
\end{acknowledgments}

\bibliography{auto_generated}
\cleardoublepage \appendix\section{The CMS Collaboration \label{app:collab}}\begin{sloppypar}\hyphenpenalty=5000\widowpenalty=500\clubpenalty=5000\vskip\cmsinstskip
\textbf{Yerevan Physics Institute, Yerevan, Armenia}\\*[0pt]
A.~Tumasyan
\vskip\cmsinstskip
\textbf{Institut f\"{u}r Hochenergiephysik, Wien, Austria}\\*[0pt]
W.~Adam, J.W.~Andrejkovic, T.~Bergauer, S.~Chatterjee, M.~Dragicevic, A.~Escalante~Del~Valle, R.~Fr\"{u}hwirth\cmsAuthorMark{1}, M.~Jeitler\cmsAuthorMark{1}, N.~Krammer, L.~Lechner, D.~Liko, I.~Mikulec, P.~Paulitsch, F.M.~Pitters, J.~Schieck\cmsAuthorMark{1}, R.~Sch\"{o}fbeck, M.~Spanring, S.~Templ, W.~Waltenberger, C.-E.~Wulz\cmsAuthorMark{1}
% ...
\vskip\cmsinstskip
% ...
1:  Also at TU Wien, Wien, Austria\\
% ...
\end{sloppypar}
\end{document}